\documentclass{article}
\usepackage{jheppub}
\usepackage{amsmath}
\usepackage{amsthm}
\usepackage{amsfonts}	
\usepackage{graphicx}
\usepackage{rotating}
\usepackage{array}
\usepackage{hyperref}

\usepackage{tikz}
\usetikzlibrary{shapes,snakes,arrows}
\usetikzlibrary{shapes.misc}
\usetikzlibrary{decorations.pathreplacing}
\usetikzlibrary{patterns}

\newcommand{\PreserveBackslash}[1]{\let\temp=\\#1\let\\=\temp}
\newcolumntype{C}[1]{>{\PreserveBackslash\centering}p{#1}}
\newcommand{\tabincell}[2]{\begin{tabular}{@{}#1@{}}#2\end{tabular}}
\title{Coulomb Branch for A-type Balanced Quivers in 3d $\mathcal{N}=4$ gauge theories}
\author[*]{Gong Cheng}
\author[**]{Amihay Hanany}
\author[*]{Yabo Li}
\author[*]{Yidi Zhao}

\affiliation[*]{Department of Modern Physics, University of Science and Technology of China, Hefei, Anhui 230026, China}

\affiliation[**]{Theoretical Physics Group, The Blackett Laboratory, Imperial College London, Prince Consort Road, London, SW7 2AZ, UK}
\emailAdd{cghope@mail.ustc.edu.cn}
\emailAdd{a.hanany@imperial.ac.uk}
\emailAdd{liyb@mail.ustc.edu.cn}
\emailAdd{zyd2117@mail.ustc.edu.cn}

\preprint{Imperial-TP-17-AH-01}
\abstract{We study Coulomb branch moduli spaces of a class of three dimensional $\mathcal{N}=4$ gauge theories whose quiver satisfies the balance condition. The Coulomb branch is described by dressed monopole operators which can be counted using the Monopole formula. We mainly focus on A-type quivers in this paper, using Hilbert Series to study their moduli spaces, and present the interesting pattern which emerges. All of these balanced A-type quiver gauge theories can be realized on brane intervals in Type IIB string theory, where mirror symmetry acts by exchanging the five branes and induces an equivalence between Coulomb branch and Higgs branch of mirror pairs.  For each theory, we explicitly discuss the gauge invariant generators on the Higgs branch and the relations they satisfy. Finally, some analysis on $D_4$ balanced quivers also presents an interesting structure of their moduli spaces. }

\begin{document}
\maketitle
\flushbottom

\section{Introduction}
An infinite class of 3d gauge theories with $\mathcal{N}=4$ supersymmetry can be realized by using the specific brane and orientifold configurations in M-theory and Type IIB string theory. They can be described by quiver diagrams which encode the essential information like gauge symmetry as well as the representation under which the fields in the theory transform. In this article, we consider the quivers with an unitary gauge group on each gauge node.	   

The balance of a  gauge node $U(N_i)$ in a simply laced ADE Series quiver is defined as \cite{Hanany:2016gbz}:

\begin{equation}
\text{Balance}ADE(i) = -2N_i +\sum \limits_{j\in \text{adjacent nodes}} N_j
\end{equation}

If all gauge nodes in a quiver have a balance of zero, the quiver is termed balanced.

We mainly focus on A-type balanced quivers and study their Coulomb branch moduli space by finding all the generators and relations. Those generators and relations are present in the form of matrices composed of quarks in the mirror theory Higgs branch. We analyze the pattern of the relations and try to make a general description. 

The moduli space of the theory can be viewed as the complex varieties described by the chiral ring of holomorphic functions. One can read the information of the moduli space from its Hilbert Series which enumerates the gauge invariant BPS operators of chiral ring. Recently, a new efficient technique was proposed in \cite{Cremonesi:2013lqa} to calculate the series, i.e. the monopole formula:

\begin{equation}
\text{HS}(t,z)=\sum\limits_{m\in \Lambda(\hat{G})/W(\hat{G})} z^{J(m)}t^{\Delta(m)}P_G(t,m)  
\end{equation} 

$\Delta(m)$ is the conformal dimension or R-charge of monopole operators, which was calculated using radial quantization in \cite{Borokhov:2002ib}. We quote the result here:
\begin{equation}\label{con}
\Delta(m)=\frac{1}{2}\sum\limits_{i=1}^n \sum\limits_{\rho_i\in R_i}|\rho_i(m)|-\sum\limits_{\alpha\in\Delta_{+}}|\alpha(m)|
\end{equation}

$m$ denotes the magnetic charge which takes value in the quotient of dual weight lattice by Weyl group. $J(m)$ is the topological charge, and is counted by fugacity $z$.  $P_G$ is the generating function of Casimir invariants which plays the role of dressing factor. The first term of conformal dimension formula (\ref{con}) comes from the hypermultiplets that transform in representation $\rho_i$. The second term accounts for the vector multiplets, and $\Delta_+$ is the set of positive roots of the gauge group. The details of monopole formula can be found in \cite{Hanany:2016ezz} and \cite{Cremonesi:2013lqa}.

It was discussed in \cite{Hanany:2016ezz} the method of dividing weight lattice into fans in order to sum up infinite terms. We found an algorithm that generalizes this method to high dimensions. This allows the computation for quivers with many gauge nodes.

The outline of this paper is as follows. In section \ref{mir} and \ref{hil}  we review the method for finding the mirror and reading the relations from Hilbert Series. In section \ref{threl} we introduce three kinds of relations of A-type balanced quiver. In section \ref{2type} we begin the analysis of $A_2$ balanced quiver and introduce the tool we adopt. In section \ref{generel} we use this method to study the general $A_n$-type balanced quiver and make prediction of some $A_4$-type quiver where the HS is difficult to calculate. Finally, we put some results of D-type balanced quiver in the section \ref{D}.

\section{Mirror symmetry}\label{mir}

The A-type quiver gauge theory can be realized in the brane configuration discussed in \cite{Hanany:1996ie}. Specifically, it is probed by observer living in the D3 brane which moves in the interval of 5-branes, and therefore is (2+1) dimensional. Each of these configurations could be described by a pair of partitions $(\rho, \sigma)$, and was denoted by $T^{\sigma}_{\rho}(G)$  in \cite{Cremonesi:2014uva}.   

At the fixed point of these SYM theories, an action called mirror symmetry \cite{Intriligator:1996ex} exchanges the Higgs branch and Coulomb branch of a mirror pair. Generally, in the 3d $\mathcal{N}=4$ gauge theory, these two branches are both hyper-K$\ddot{a}$hler spaces. The Higgs branch is protected from quantum correction and therefore the classical description is enough, while Coulomb branch does receive quantum correction and is described by dressed monopole operators.

Mirror symmetry acts on brane configuration by exchanging D5 brane and NS5 brane, with their linking number also exchanged. The linking number is conserved quantity of each 5-brane in brane transitions. The definition is:

\begin{equation}
\begin{split}
L_{NS}=\frac{1}{2}(r-l)+(L-R)\\
L_{D}=\frac{1}{2}(r-l)+(L-R)
\end{split}
\end{equation}

For each NS5 (D5) brane, $r$ denotes the number of D5 (NS5) branes to the right of this NS5 (D5) brane, $l$ denotes the number of D5 (NS5) branes to the left of NS5 (D5) brane. The second term is the net number of D3 brane ending on each 5-brane, from left minus from right.

When all the D5 branes are to the one side of all the NS5 branes, we only need to count for the net number of D3 branes to keep track of the linking numbers.  In these configuration, ($\sigma_1$, $\sigma_2$,\dots,$\sigma_l$) corresponds to the net number of D3 branes on each D5 brane from interior to exterior, and ($\rho_1$, $\rho_2$, \dots ,$\rho_{l'}$ ) corresponds to the net number of D3 branes on each NS5 brane from interior to exterior.  Given the data of $\rho$ and $\sigma$, one can recover the quiver diagram in the following way:

Suppose that $\rho=(\rho_1,\ldots,\rho_{l^\prime})$ and $\sigma=(\sigma_1,\ldots,\sigma_l)$ are two partitions of $N$, which satisfies:
\begin{equation}
\sigma_1\geq\ldots\geq\sigma_l>0,\quad \rho_1\geq\ldots\geq\rho_{l^\prime}>0,\quad \sum_{i=1}^l\sigma_i=\sum_{i=1}^{l^\prime}\rho_i=N
\end{equation}
The quiver diagram for $T_\rho^\sigma(SU(N))$ is as follows:
\begin{figure}[htbp]
	\centering
		\begin{tikzpicture}
  \node (g1) at (0,0) [circle,draw,label=above:{$N_1$}] {};
	\node (g2) at (1,0) [circle,draw,label=above:{$N_2$}] {};
	\node (g3) at (3.5,0) [circle,draw,label=above:{$N_{l^\prime-1}$}] {};
	\node (f1) at (0,-1) [regular polygon,regular polygon sides=4,draw,,label=below:{$M_1$}] {};
	\node (f2) at (1,-1) [regular polygon,regular polygon sides=4,draw,,label=below:{$M_2$}] {};
	\node (f3) at (3.5,-1) [regular polygon,regular polygon sides=4,draw,,label=below:{$M_{l^\prime-1}$}] {};
	\draw (g1)--(g2) (g1)--(f1) (g2)--(f2) (g3)--(f3);
	\draw [dashed] (g2)--(g3);
\end{tikzpicture}
		\caption{General A-type Quiver}
\end{figure}
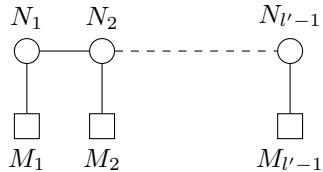

$l$ and $l'$ are the length of partition $\sigma$ and $\rho$, $\hat{l}$ is the length of transpose of $\sigma$.
\begin{equation}
\begin{split}
M_j=&\sigma^T_j-\sigma^T_{j+1},\ \text{with}\: \sigma^T_i=0\:\text{for}\: i\geq \hat{l}+1\\
&N_j=\sum_{k=j+1}^{l^\prime}\rho_k-\sum_{i=j+1}^{\hat{l}}\sigma^T_i
\end{split}
\end{equation}

We can get $\rho$ and $\sigma$ using the above equations. Then finding the mirror can easily be done by exchanging $\rho$ and $\sigma$, corresponding to exchanging of the linking number of 5-branes.

\section{Hilbert Series and Plethystic Logarithm}\label{hil}
The Hilbert Series (HS) is a partition function counting BPS operators in the chiral ring according to their charges in different symmetry groups. For example, a (dressed) monopole operator is characterized by its topological charge $J(m)$ and conformal dimension $\Delta(m)$, and appears in HS in the form of $z^{J(m)}t^{\Delta(m)}$. The generators of the chiral ring can be conveniently read from the Plethystic Logarithm in \cite{Benvenuti:2006qr}. 

The Plethystic Logarithm (PL) is the reverse operation of Plethystic Exponential, which generates the symmetric products of different orders. It can be calculated iteratively if we know the expansion of Hilbert Series.
	
For example, the unrefined Hilbert Series of quiver (1)-(2)-[3] is 

\begin{equation}
\text{HS}_{(1)-(2)-[3]}=1+8t+35t^2+111t^3+...
\end{equation}

Then we try to find whose symmetric products give this expression. From the first order term, we know that $\text{PL}=8t+O(t^2)$, but $sym^2[8t]=36t^2$. So we need a term $ -t^2$ in the PL to compensate and get the correct $35t^2$ in the HS.  So we know that $\text{PL}=8t-t^2+O(t^3)$. One can do this order by order and finally gets $\text{PL}=8t-t^2-t^3$. 

The refined version of this process gives the Plethystic Logarithm of refined Hilbert Series as:

\begin{equation}
\text{PL}=(2+z_1+z_2+\frac{1}{z_1}+\frac{1}{z_2}+z_1z_2+\frac{1}{z_1z_2})t-t^2-t^3
\end{equation}
  
The first several positive terms are generators. Their topological charges constitute the weights of adjoint representation of SU(3) group, the enhanced topological symmetry in the original theory and flavour symmetry in the mirror theory perspective. The weights are measured by the basis of simple roots, i.e. $z_1$ and $z_2$ correspond to the two simple roots of SU(3). Therefore, in order to write it in the standard form of SU(3) characters, one need to do a transformation using Cartan matrix in \cite{Cremonesi:2013lqa} to shift the basis from simple roots to fundamental weights in Weyl chamber. In our subsequent results of Hilbert Series or PL, we will denote the character using Dynkin label. So in this example, it is represented as $\text{PL}=[11]t-t^2-t^3$. 

 The negative terms represent relations, and they transform in singlet representation. These information helps people finding the concrete form of relations. 

In summary, we first calculate the Hilbert Series using monopole formula, and then take Plethystic Logarithm of it and learn how the relations transform under the enhanced symmetry. Finally, we come to the Higgs branch of the mirror theory, which do not receive quantum correction, and make use of F-term to find the concrete form of relations.

\section{Three kinds of relations}\label{threl}
\begin{figure}[h]%
\centering
\begin{tikzpicture}
	\node (g1) at (-2,0) [circle,draw,label=below:1] {};
	\node (g2) at (-1,0) [circle,draw,label=below:2] {};
	\node (g3) at (0,0) [circle,draw,label=below:3] {};
	\node (g4) at (1,0)[circle,draw,label=below:2] {};
	\node (g5) at (2,0)[circle,draw,label=below:1] {};
	\node (f1) at (0,1)[regular polygon,regular polygon sides=4,draw,label=above:4] {};
	\draw (g1)--(g2) --(g3)--(g4)--(g5)
				(f1)--(g3);
	\node at (-0.5,0.2) {$R_1$};
	\node at (-0.5,-0.2) {$L_1$};
	\node at (0.5,0.2) {$r_1$};
	\node at (0.5,-0.2) {$l_1$};
	\node at (1.5,0.2) {$r_2$};
	\node at (1.5,-0.2) {$l_2$};
	\node at (-0.2,0.5) {$u$};
	\node at (0.2,0.5) {$d$};
\end{tikzpicture}
\caption{Mirror quiver 1-2-3-2-1}%
\label{}%
\end{figure}
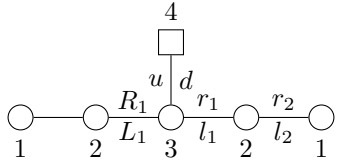%
A general A-type quiver satisfies three kinds of relations, i.e. matrix relation, trace relation and meson relation. The first two types of relations, matrix and trace relations, only rely on the shape of mirror quiver.

Suppose we have $k$ generators:
\begin{equation}
M_1=du,\ \ \ \
M_2=dr_1l_1u,\ \ \
\dots\dots \ \ \
,\ M_k=dr_1r_2\dots l_2l_1u
\end{equation}

The relations they satisfy can be read from F-terms:
\begin{equation}
\begin{split}
ud+r_1l_1+L_1R_1=0\\
r_2l_2-l_1r_1=0\\
l_2r_2=0\\
R_1L_1-L_2R_2=0\\
R_2L_2=0
\end{split}
\end{equation}
$r$ and $l$ are the edges of short arm of mirror quiver, $R$ and $L$ are those of long arm of mirror quiver (as shown in the above figure).
Using the F-term relations, we can turn all the $r_il_i$ to $r_1l_1$. For example, $dr_1r_2l_2l_1u=dr_1l_1r_1l_1u$. 
In the following several parts, we will discuss these three kinds of relations separately.
\subsection{Trace relation}
\textit{Proposition}: the $N$th order trace relation is 
\begin{equation}
\text{Tr} (\Sigma _{\lambda }\frac{1}{m_{\lambda }}M_{i_1} M_{i_2}\dots M_{i_r})=0
\end{equation}

We denote $\lambda' =\{i'_1,i'_2\text{...}i'_r\}$ as all the possible partition of $N$ (not ordered) that satisfies $i'_1+i'_2+\dots+i'_r=N$, with $i'_s\leq k$, where $k$ is the number of generators. Then  $\lambda =\{i_1,i_2\text{...}i_r\}$ is the quotient of $\lambda'$ by cyclic group. e.g.  $\lambda'=\{1,2,3\}$ is identified with $\lambda'=\{2,3,1\}$ to give $\lambda=\{1,2,3\}$.   $m_{\lambda}$ is the power of monomial $M_{i_1} M_{i_2}\dots M_{i_r}$ when $M_{i_1} M_{i_2}\dots M_{i_r}$ can be written as $(M_{i_1} M_{i_2}\dots M_{i_s})^{m_{\lambda}}$, for some $s<r$.

This relation was conjectured in \cite{Hanany:2011db}.

To see an example, consider the 6th order trace relation with 2 generators. It is 
\begin{equation}
\text{Tr}(\frac{1}{6}{M_1}^6+\frac{1}{3}{M_2}^3+{M_2}^2{M_1}^2+\frac{1}{2}(M_2M_1)^2+M_2{M_1}^4)=0
\end{equation}

The proof of the proposition is as follows. 

\textit{Proof}:
Degree $i$ generator:
\begin{equation}
M_i=d_1 r_1...r_il_i...l_1u_1=d_1(r_1l_1)^{i-1}u_1
\end{equation}

We consider trace of product of generators in degree $N$:
\begin{equation}
\begin{split}
&\text{Tr}(M_iM_jM_k...)\\=&\text{Tr}((r_1l_1)^{i-1}u_1d_1(r_1l_1)^{j-1}u_1d_1(r_1l_1)^{k-1}u_1d_1...u_1d_1)\\=&\text{Tr}((r_1l_1)^{i-1}(-L_1R_1-r_1l_1)(r_1l_1)^{j-1}(-L_1R_1-r_1l_1)(r_1l_1)^{k-1}\\&(-L_1R_1-r_1l_1)...(-L_1R_1-r_1l_1))
\end{split}
\end{equation}

That is to say,
\begin{equation}
\text{Tr}(M...)=\text{Tr}(\text{combinations of}\ (-L_1R_1-r_1l_1)\ \text{and}\ r_1l_1)
\end{equation}

We add up all the possible degree $N$ terms of this form, and name it $S$:
\begin{equation}
S=\text{Tr}(r_1l_1...r_1l_1)+...+\text{Tr}((-L_1R_1-r_1l_1)...(-L_1R_1-r_1l_1))
\end{equation}

$S$ can also be written by combination of traces of products of generators:
\begin{equation}
\begin{split}
S=&N\text{Tr}(M_N)+N\sum_{1\le s_1...\le s_m\le N-1}\delta(N-\sum_{i=1}^{m}s_i)\widehat{\sum_{\sigma\in S_m}}\frac{\text{Tr}(\prod_{i=1}^{m}M_{s_\sigma(i)})}{\mathcal{M}(\text{Tr}(\prod_{i=1}^{m}M_{s_\sigma(i)}))}\\&+\sum_{1\le r\le \frac{N}{2},r|N}r\sum_{1\le s_1...\le s_m\le r}\delta(r-\sum_{j=1}^{l}s_j)\widehat{\sum_{\rho\in S_l}}\frac{\text{Tr}[(\prod_{j=1}^{l}M_{s_\rho(j)})^{N/r}]}{\mathcal{M}(\text{Tr}(\prod_{j=1}^{l}M_{s_\rho(j)})^{N/r})}\\&+\text{Tr}(r_1l_1r_1l_1...r_1l_1)
\end{split}
	\end{equation}
where $\widehat{\sum}_{\sigma\in S_m}$ denotes the summation over $\sigma\in S_m$ such that $\prod_{i=1}^{m}M_{s_\sigma(i)}$ is not an operator with an integer power greater than 1, and $\mathcal{M}(\text{Tr}(\prod_{i=1}^{m}M_{s_\sigma(i)}))$ denotes the multiplicity of $\text{Tr}(\prod_{i=1}^{m}M_{s_\sigma(i)})$.

According to the F-terms, the last term equals to 0. And now we are able to prove that $S$ equals to 0.

$S$ is the trace of the sum of all possible terms combining $(-L_1R_1-r_1l_1)$ and $r_1l_1$. We expand the parenthesis, then it's a polynomial of $L_1R_1$ and $r_1l_1$.

We first consider terms without $L_1R_1$. According to F-terms, these terms equal to 0.

Now we consider terms with one $L_1R_1$. All these terms are in same configuration, because of the rotation invariance of trace. (\# of rotations of the configuration) = $N$.
\begin{equation}
\begin{split}
(\#\ \text{of terms})&=(\#\ \text{of rotations of the configuration})\cdot\sum_{k=1}^{N}(-1)^{k-1}C_{N-1}^{k-1}\\&=N(1-1)^{N-1}=0
\end{split}
\end{equation}

Then we consider terms with two consecutive $L_1R_1$. In this case we also have (\# of rotations of the configuration) = $N$.
\begin{equation}
\begin{split}
(\#\ \text{of terms})&=(\#\ \text{of rotations of the configuration})\cdot\sum_{k=2}^{N}(-1)^{k-2}C_{N-2}^{k-2}\\&=N(1-1)^{N-2}=0
\end{split}
\end{equation}

Similarly, we can compute the number of terms in arbitrary configuration(with $n \ L_1R_1$), and find out it's 0:
\begin{equation}
\begin{split}
(\#\ \text{of terms})&=(\#\ \text{of rotations of the configuration})\cdot\sum_{k=n}^{N}(-1)^{k-n}C_{N-n}^{k-n}\\&=(\#\ \text{of rotations of the configuration})\cdot(1-1)^{N-n}=0
\end{split}
\end{equation}

Finally, there are some terms which are purely the products of $N$ $L_1R_1$, but as our first case, they equal to 0 because of the F-terms.

Now we know that each term of $S$ equals to 0, then $S$ = 0. Therefore we conclude that:
\begin{equation}
\begin{split}
0=&N\text{Tr}(M_N)+N\sum_{1\le s_1...\le s_m\le N-1}\delta(N-\sum_{i=1}^{m}s_i)\widehat{\sum_{\sigma\in S_m}}\frac{\text{Tr}(\prod_{i=1}^{m}M_{s_\sigma(i)})}{\mathcal{M}(\text{Tr}(\prod_{i=1}^{m}M_{s_\sigma(i)}))}\\&+\sum_{1\le r\le \frac{N}{2},r|N}r\sum_{1\le s_1...\le s_m\le r}\delta(r-\sum_{j=1}^{l}s_j)\widehat{\sum_{\rho\in S_l}}\frac{\text{Tr}[(\prod_{j=1}^{l}M_{s_\rho(j)})^{N/r}]}{\mathcal{M}(\text{Tr}(\prod_{j=1}^{l}M_{s_\rho(j)})^{N/r})}
\end{split}
\end{equation}

\mbox{}

Let's have a quick check for the simplest case, $N$=2.
\begin{equation}
\begin{split}
&\text{Tr}(r_1l_1(-L_1R_1-r_1l_1)+(-L_1R_1-r_1l_1)r_1l_1+(-L_1R_1-r_1l_1)(-L_1R_1\\&-r_1l_1)+r_1l_1r_1l_1)\\=&\text{Tr}(-L_1R_1-L_1R_1+0\cdot r_1l_1r_1l_1-0\cdot L_1R_1r_1l_1)=0
\end{split}
\end{equation}

And,
\begin{equation}
\begin{split}
\text{Tr}(M_2)&=\text{Tr}(r_1l_1(-L_1R_1-r_1l_1))\\
\text{Tr}({M_1}^2)&=\text{Tr}((-L_1R_1-r_1l_1)(-L_1R_1-r_1l_1))
\end{split}
\end{equation}

Therefore we have:
\begin{equation}
2\text{Tr}(M_2)+\text{Tr}({M_1}^2)=0
\end{equation}

\subsection{Matrix relation}
The $N$th order matrix relation is 
\begin{equation}
\Sigma _{\lambda' }  M_{i_1'} M_{i_2'}\dots M_{i'_r}=0
\end{equation}

Again $\lambda' =\{i'_1,i'_2\text{...}i'_r\}$ is one of all the possible partitions of $N$ (not ordered), with $i_s\leq k$. This time we do not quotient it by the cyclic group.

\textit{Proposition}: Suppose the length of the long arm of mirror quiver is $l$, there will be a matrix relation of the above form in each order starting from $(l+1)$th order.

\textit{Proof}: Suppose we have $k$ generators, and we want to prove $n$th $(n>k)$ order matrix relation. 

First we enlarge the generators $\{M_1,M_2,\dots,M_k\}$ to a larger set $\{M_1,M_2,\dots,M_n\}$. For $s>k$, let 
\begin{equation}
M_s=dr_1l_1 r_1l_1... r_1l_1 u=0.
\end{equation}

Note that the product of $r_1l_1$'s, in which the number of $r_1l_1$ exceeds the length of short arm of mirror quiver, is zero.
We claim that: 
\begin{equation}
\Sigma _{\lambda' } M_{i'_1} M_{i'_2}\dots M_{i'_r}=(-1)^{n-1}dL_1R_1 L_1R_1 \ldots L_1R_1 u
\end{equation}
where $\lambda' =\{i'_1,i'_2\text{...}i'_r\}$ is a partition of $n$, with $i_r\leq n$.

Observe that 
\begin{equation}
\Sigma _{\lambda' } M_{i'_1} M_{i'_2}\dots M_{i'_r}=d(\text{combinations of}\ r_1l_1\ \text{and}\ -(r_1l_1+L_1R_1))u
\end{equation}

For example
\begin{equation}
\begin{split}
&{M_1}^4+M_2 {M_1}^2+ M_1M_2M_1+{M_1}^2M_2+M_3 M_1+M_1M_3+{M_2}^2+M_4\\
=&d(r_1l_1 r_1l_1 r_1l_1-(r_1l_1+L_1R_1) r_1l_1 r_1l_1-r_1l_1(r_1l_1+L_1R_1)r_1l_1-r_1l_1 r_1l_1(r_1l_1\\&+L_1R_1)+(r_1l_1+L_1R_1) (r_1l_1+L_1R_1) r_1l_1+(r_1l_1+L_1R_1) r_1l_1 (r_1l_1+L_1R_1)\\&+r_1l_1(r_1l_1+L_1R_1) (r_1l_1+L_1R_1)-(r_1l_1+L_1R_1) (r_1l_1+L_1R_1)(r_1l_1+L_1R_1))u
\end{split}
\end{equation}

There should be $2^n$ terms. The strategy is to find $2^{n-1}$ pairs and pair them up to turn the rightmost term to $L_1R_1$: 
\begin{equation}
\begin{split}
&d(-r_1l_1 r_1l_1+r_1l_1 (r_1l_1+L_1R_1)-(r_1l_1+L_1R_1) (r_1l_1+L_1R_1)
+(r_1l_1\\&+L_1R_1)r_1l_1)L_1R_1u
\end{split}
\end{equation}

Then repeat the process to find $2^{n-2}$ pairs and pair them up to get:
\begin{equation}
d( r_1l_1-(r_1l_1+L_1R_1) )L_1R_1L_1R_1u
\end{equation}
and finally get:
\begin{equation}
-d(L_1R_1 L_1R_1 L_1R_1)u
\end{equation}

When the number of $L_1R_1$'s exceeds the length of long arm of mirror quiver, the term $dL_1R_1L_1R_1...L_1R_1u$ vanishes. This finishes the proof of the proposition.

\subsection{Meson relation}

In general, meson relation is referred to as the relation which involves the anti-symmetrization of upper and lower indices separately. For example, for $A_3$ the second meson relation $(M_1)_{[i}^{\phantom{k]}[j}(M_2)_{k]}^{\phantom{k]}l]}$ transforms in $[010]\times[010]=[020]+[101]+1$. For $A_4$, it transforms in $[0100]\times[0010]=[0110]+[1001]+1$. In general, the two indices of $M$ are in the [10...0] and [0...01] respectively. Thus, the $k$th anti-symmetrization of lower indices give $[\underbrace{00\ldots1}_k\ldots0]$. The $k$th anti-symmetrization of up indices is $[0\ldots\underbrace{1\ldots00}_k]$. 

If $M_1=M_2=M$, $(M_1)_{[i}^{\phantom{k]}[j}(M_2)_{k]}^{\phantom{k]}l]}=M_{[i}^{\phantom{k]}[j}M_{k]}^{\phantom{k]}l]}$ is just the 2 by 2 minor of $M$, and its vanishing means $\text{rank}(M)\leq 1$. 
We take $A_3$-type quiver [2]-(2)-(2)-(2)-[2] as an example. Its mirror has the following form.
\begin{figure}[h]%
\centering
\begin{tikzpicture}
	\node (g1) at (-1,0) [circle,draw,label=below:1] {};
	\node (g2) at (0,0) [circle,draw,label=below:2] {};
	\node (g3) at (1,0) [circle,draw,label=below:1] {};
	\node (f1) at (0,1) [regular polygon,regular polygon sides=4,draw,label=above:4] {};
	\draw (g1)--(g2)--(g3) (g2)--(f1);
	\node at (0.5,0.3) {$r_1$};
	\node at (0.5,-0.3) {$l_1$};
	\node at (-0.2,0.5) {$u$};
	\node at (0.2,0.5) {$d$};
\end{tikzpicture}
\caption{Quiver 1-2-1}%
\label{}%
\end{figure}
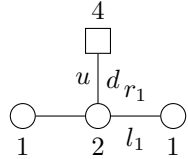

The generators are $M_1=du$ and $M_2=d r_1 l_1 u$. 

\textit{Proposition}:  $(M_1)_{[i}{}^{[j}(M_2)_{k]}{}^{l]}=0$

\textit{Proof}:
$$(M_1)_i{}^j=d_i{}^\alpha u_\beta {}^j \delta_\alpha {}^\beta,i,j=1,2,3,4;\alpha,\beta=1,2;$$
$$(M_2)_k{}^l=d_k{}^\gamma u_\kappa{}^l Q_\gamma{}^\kappa,\text{where}\ Q_\gamma{}^\kappa=(r_1l_1)_\gamma{}^\kappa=(r_1)_\gamma{}^1(l_1)_1{}^\kappa$$
So
\begin{equation}
\begin{split}  
(M_1)_{[i}{}^{[j}(M_2)_{k]}{}^{l]} &= d_{[i}{}^\alpha d_{k]}{}^\gamma u_\beta {}^{[j} u_\kappa{}^{l]} \delta_\alpha {}^\beta Q_\gamma{}^\kappa\\
&=d_i{}^{[\alpha} d_k{}^{\gamma]} u_{[\beta} {}^j u_{\kappa]}{}^l \delta_\alpha {}^\beta Q_\gamma{}^\kappa\\
&=d_i{}^\alpha d_k{}^\gamma u_\beta {}^j u_\kappa{}^l \delta_{[\alpha} {}^{[\beta} Q_{\gamma]}{}^{\kappa]}
\end{split}
\end{equation}
and 
\begin{equation}
\begin{split}
\delta_{[\alpha} {}^{[\beta} Q_{\gamma]}{}^{\kappa]}&=\epsilon_{\beta\kappa}\epsilon^{\alpha\gamma}\delta_\alpha {}^\beta Q_\gamma{}^\kappa\\
&=\delta_\kappa{}^\gamma  Q_\gamma{}^\kappa\\
&=\text{Tr}(Q)
\end{split}
\end{equation}

Using F-term, we have $\text{Tr}(Q)=\text{Tr}(r_1l_1)=\text{Tr}(l_1r_1)=0$. So $(M_1)_{[i}{}^{[j}(M_2)_{k]}{}^{l]}=0$.
Generator $M_2$ is of rank 1, so $(M_2)_{[i}{}^{[j}(M_2)_{k]}{}^{l]}=0$. 
In sum, there are two meson relations: $(M_1)_{[i}{}^{[j}(M_2)_{k]}{}^{l]}=0$ and $(M_2)_{[i}{}^{[j}(M_2)_{k]}{}^{l]}=0$.

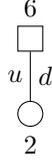
\begin{figure}[h]%
\centering
\begin{tikzpicture}
	\tikzstyle{gauge}=[circle,draw];
	\tikzstyle{flavor}=[regular polygon,regular polygon sides=4,draw];
	\node (g1) at (0,0) [gauge,label=below:2] {};
	\node (f1) at (0,1) [flavor,label=above:6] {};
	\draw (g1)--(f1);
	\node at (-0.2,0.5) {$u$};
	\node at (0.2,0.5) {$d$}; 
\end{tikzpicture}
\caption{Quiver 2-[6]}%
\label{aa}
\end{figure}

In addition to this kind of meson relation with the anti-symmetrization of two matrices, other meson relations involving more matrices also exist. For example, we consider quiver \{\{12221\},\{01010\}\}\footnote[1]{Meaning of this symbol: the first parenthesis contains the gauge nodes, and the second parenthesis contains the flavour nodes at the corresponding position.} whose mirror is Figure \ref{aa}

The generator $M$ has rank 2, so one would expect a relation $M_{[i}{}^{[j}{M_k}^lM_{p]}{}^{q]}=0$. It transforms in $\Lambda^3[10000]\times \Lambda^3[00001]=[00200]+[01010]+[10001]+[00000]$. This relation  emerges at $A_5$ quiver, because for lower rank quiver like $A_4$, $\Lambda^3[1000]\times\Lambda^3[0001]=[0010]\times[0100]=\Lambda^2[1000]\times\Lambda^2[0001]$. In general, for every $A_n$-type quiver with $n$ odd, a meson that transforms in a new representation emerges. 

For higher degree generators, the meson relation can be proved iteratively (just as we get rid of outermost $d$ and $u$ and leave the anti-symmetrized inner part, one can continue to do this to get rid of the outer part).

\section{$A_2$-type quiver}\label{2type}
\subsection{General pattern}
The $A_2$-type quiver has a very clear pattern.\footnote[2]{See Appendix \ref{some} for the PL and relations of some $A_2$-type quivers.} First, from the Hilbert Series, one conjectures that there are only trace relations (in singlet representation) and matrix relations (in adjoint representation). Indeed, we have given all the matrix and trace relations by the proposition proved above.
By observation one can extract the pattern which is satisfied for general $A_2$ balanced quiver.
In the following we denote the quiver by $\{k_1,k_2\}$ ($k_1 \leq k_2$, they are numbers in gauge nodes).

Pattern:
\begin{itemize}
\item The matrix relation starts at order $k_2+1$. 
\item The matrix relation terminates at order $2k_1$.
\item The trace relation exists from first order to order $k_1+k_2-1$ for non-complete intersection, and from first order to order $3k_1$ for complete intersection. 
\end{itemize}
The first rule can be explained by the proof of matrix relation above. Recall that 
\begin{equation}
\Sigma _{\lambda' } M_{i'_1} M_{i'_2}\dots M_{i'_r}=(-1)^{n-1}d\underbrace{L_1R_1 L_1R_1 \ldots L_1R_1}_{n-1} u
\end{equation}

The right hand side is 0 when $n\geq k_2+1$. So the matrix relation starts at order $k_2+1$.

However, the second and third properties take some efforts to explain. It's clear that the higher order matrix relation doesn't show up because it is not independent from the previous ones. We will give the detailed explanation later.
\subsection{Why no meson relation?}
The adjoint representation in [1,1] can also be realized by second meson relation, like 
\begin{equation}
(M_1)_{[i}{}^{[j}(M_2)_{k]}{}^{l]}=0
\end{equation}
besides matrix relation. But we conjecture that it is equivalent 
to the matrix relation. Here is the argument:
first, we show that the meson relation can always be contracted to give a matrix relation. For example,
the second meson relation $(M_1)_{[i}{}^{[j}(M_2)_{k]}{}^{l]}=0$ can be contracted to:
\begin{equation}
M_1 M_2+M_2 M_1-\text{tr}(M_1)M_2-\text{tr}(M_2)M_1=0
\end{equation}

Subsequently, to avoid ambiguity, we will refer to the matrix relation proved in the proposition as the general matrix relation to distinguish from those gotten from contraction.
It's clear that the number of d.o.f. of the contracted meson relation is the same as that of the original meson relation, as they are in the same representation [1,1]. Moreover, contracting some indices of a tensor gives a linear combination of its components. Therefore, we don't lose information after doing this contraction if the number of d.o.f. doesn't decrease.  So if we want to prove that certain meson relation is equivalent to a matrix relation, we only need to contract the meson relation, and then prove the contracted relation is equivalent to the matrix relation. We see some examples in the following.

\subsection{Examples}

\subsubsection{Quiver [2]-(2)-(2)-[2]}
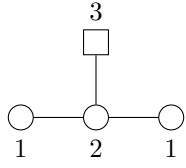
\begin{figure}[h]
	\centering
	\begin{tikzpicture}
	\node (g1) at (-1,0) [circle,draw,label=below:1] {};
	\node (g2) at (0,0) [circle,draw,label=below:2] {};
	\node (g3) at (1,0) [circle,draw,label=below:1] {};
	\node (f1) at (0,1) [regular polygon,regular polygon sides=4,draw,label=above:3] {};
	\draw (g1)--(g2)--(g3) (g2)--(f1);
\end{tikzpicture}
	\caption{Mirror 1-2-1}
	\label{fig:fig4}
\end{figure}

By the brane construction, the mirror quiver is: see figure \ref{fig:fig4}. Now we focus on the Coulomb branch of [2]-(2)-(2)-[2] which corresponds to the Higgs branch of the mirror theory.

\textbf{Plethystic Logarithm}
\begin{equation}
\text{PL}=[11]t+[11]t^2-([11]+2)t^3-([11]+1)t^4+\dots
\end{equation}

The generators with spin 1, spin 2  are in the adjoint representation of SU(3). 
  
\textbf{Relations}

From the PL one can write down the relations according to the representations that they belong to. 

3rd order meson relation
\begin{equation}\label{me}
(M_1)_{[i}{}^{[j}(M_2)_{k]}{}^{l]}=0
\end{equation}
(For simplicity, we denote this by $M_1**M_2=0$.)

And

\begin{equation}
(M_1)_{[i}{}^{[j}(M_1)_k{}^l(M_1)_{p]}{}^{q]}=\text{Det}(M_1)=0\label{dme}
\end{equation}

After contraction, Eq.(\ref{me}) becomes:
\begin{equation}
-\text{Tr}(M_2) M_1+M_1 M_2+M_2 M_1=0\label{cme}
\end{equation}

Note that after contraction we have the same number of d.o.f as in Eq.(\ref{me}), so they are equivalent. 

3rd order general matrix relation
\begin{equation}\label{any}
{M_1}^3+M_1M_2+M_2M_1=0
\end{equation}

3rd order general trace relation
\begin{equation}
\text{Tr}(\frac{1}{3}{M_1}^3+M_1M_2)=0
\end{equation}

We claim that the 3rd order meson relations (Eq.(\ref{cme}) and Eq.(\ref{dme})) can be derived from the 3rd order general matrix relation and trace relation. Details can be found in Appendix \ref{2222}.

4th order meson relation
\begin{equation}
M_2**M_2=0
\end{equation}

4th order general matrix relation
\begin{equation}\label{4lgm}
{M_2}^2+{M_1}^4+M_2{M_1}^2+M_1M_2M_1+{M_1}^2M_2=0
\end{equation}

The 4th order meson relation $M_2**M_2=0$  is equivalent to the matrix relation (\ref{4lgm}). The proof can be found in Appendix \ref{2222}.

\subsubsection{Quiver [3]-(3)-(3)-[3]}

\begin{figure}[h]
	\centering
		\begin{tikzpicture}
	\node (g1) at (-2,0) [circle,draw,label=below:1] {};
	\node (g2) at (-1,0) [circle,draw,label=below:2] {};
	\node (g3) at (0,0) [circle,draw,label=below:3] {};
	\node (g4) at (1,0)[circle,draw,label=below:2] {};
	\node (g5) at (2,0)[circle,draw,label=below:1] {};
	\node (f1) at (0,1)[regular polygon,regular polygon sides=4,draw,label=above:3] {};
	\draw (g1)--(g2) --(g3)--(g4)--(g5)
				(f1)--(g3);
\end{tikzpicture}
	\caption{Mirror 1-2-3-2-1}
	\label{fig:fig5}
\end{figure}
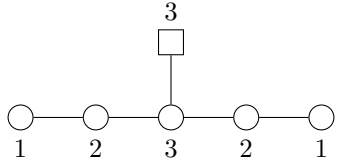

By brane construction, the mirror quiver is: see figure \ref{fig:fig5}. 

We focus on Coulomb branch of quiver [3]-(3)-(3)-[3], which corresponds to the Higgs branch of the mirror theory.

\textbf{Plethystic Logarithm}
\begin{equation}
\text{PL}=[11]t+[11]t^2+[11]t^3-([11]+2)t^4-([11]+2)t^5-([11]+1)t^6+\dots
\end{equation}

The three generators are of spin 1, spin 2, spin 3. They are all in the adjoint representation of SU(3). 

\textbf{Relations}
	
4th order meson relation
\begin{equation}\label{mes}
M_1**M_1**M_2=0
\end{equation}

(See Appendix \ref{3333} for the proof.)

Contracting it for three times
\begin{equation}\label{ct}
-\text{Tr}({M_1}^2)\text{Tr}(M_2)+2\text{Tr}( {M_1}^2M_2)=0
\end{equation}

Note that the original meson relation $M_1**M_1**M_2=0$ just has one d.o.f. since the matrices are of $3\times3$. So relation (\ref{mes}) is equivalent to its contraction.

4th order general matrix relation
\begin{equation}\label{4gm}
{M_1}^4+{M_2}^2+(M_1M_3+M_3M_1)+(M_2{M_1}^2+M_1M_2M_1+{M_1}^2M_2)=0
\end{equation}

4th order general trace relation
\begin{equation}\label{r}
\text{Tr}(\frac{1}{4}{M_1}^4+\frac{1}{2}{M_2}^2+{M_1}^2M_2+M_1M_3)=0
\end{equation}

One can check that the general matrix and trace relations can be combined to deduce the relation (\ref{ct}), and hence the meson relation (\ref{mes}).

We can utilize (\ref{mes}) to extract more information for later use:

Contract it twice
\begin{equation}
-2 \text{Tr}({M_1}^2)M_2-4 \text{Tr}(M_1 M_2)M_1+4 (M_2 {M_1}^2+ {M_1}^2M_2+M_1 M_2 M_1)-4 {M_1}^2 
\text{Tr}M_2=0
\end{equation}

Combine this with the general matrix relation (\ref{4gm}), one gets
\begin{equation}\label{dr}
\frac{1}{2}M_2**M_2+M_1**M_3=0
\end{equation}

This will be useful when we consider the 6th order relation.

\mbox{}

5th order meson relations 
\begin{equation}
\begin{split}
M_1**M_1**M_3=0\\ 
M_2**M_2**M_1=0\\ 
M_2**M_3=0  
\end{split}
\end{equation}

5th order general trace relation
\begin{equation}\label{gtt}
\text{Tr}(\frac{1}{5}{M_1}^5+M_2{M_1}^3+{M_2}^2M_1+M_2M_3+M_3{M_1}^2)=0
\end{equation}

 5th order general matrix relation  
\begin{equation}\label{gm}
\begin{split}
{M_1}^5+(M_2{M_1}^3+permu.&)+({M_2}^2M_1+permu.)+(M_3M_2+M_2M_3)\\
+(M_3{M_1}^2+permu.)=0
\end{split}
\end{equation}

$permu.$ means the terms consist of all the possible permutations of $M_1$ and $M_2$, e.g.
\begin{equation}
 ({M_1}^2M_2+permu.) = {M_1}^2M_2+M_1M_2M_1+M_2{M_1}^2
\end{equation}

In Eq.(\ref{gm}), terms in the first parenthesis all have three $M_1$ multiplied with one $M_2$, which are contained in the contraction of $M_1**M_1**M_1**M_2$:
\begin{equation}\label{long}
\begin{split}
&-({M_1}^3M_2+permu.)+\frac{1}{3}\text{Tr}({M_1}^3)M_2+\text{Tr}({M_1}^2M_2)M_1+\text{Tr}(M_1M_2){M_1}^2\\&+\frac{1}{2}(M_1M_2+M_2M_1)\text{Tr}({M_1}^2)
-\frac{1}{2}\text{Tr}({M
_1}^2)\text{Tr}(M_2)M_1+(\text{Tr}M_2){M_1}^3=0
\end{split}
\end{equation}

This expression equals zero trivially, since the matrices here are all $3\times3$ and the mesons consisting of more than three matrices must vanish. From this equation, one can solve for $({M_1}^3M_2+permu.)$ in terms of more tractable terms. Similarly, the permutations of ${M_2}^2M_1$ in the second parenthesis are contained in the contraction of $M_2**M_2**M_1$. Terms in the next parenthesis are contained in contraction of $M_2**M_3$. Finally the permutations of $M_3M_1^2$ are contained in the contraction of $M_3**M_1**M_1$. After doing all the contractions and adding these four parts together, one gets the left hand side of the 5th order general matrix relation (\ref{gm}). 

One can first prove that $M_2**M_2**M_1=0$ and $M_3**M_1**M_1=0$ hold as a result of trace relations of this order. Then combine these with the above argument,  it is easy to see that the 5th order matrix relation implies the meson relation $M_2**M_3=0$. For the detail of the proof, see Appendix \ref{3333}.

\mbox{}

6th order meson relations
\begin{equation}
\begin{split}
M_3**M_2**M_1=0\\
M_3**M_3=0\\
M_2**M_2**M_2=0
\end{split}
\end{equation}

Note that $M_3**M_2**M_1=0$ is not an independent relation, since it can be derived from $M_3**M_2=0$. $M_2**M_2**M_2=0$ is a result of relation (\ref{dr}), because one can `star' a $M_2$ to this relation and get
\begin{equation}
\frac{1}{2}M_2**M_2**M_2+M_2**M_1**M_3=0
\end{equation}

But the second term is zero from $M_2**M_3=0$. So $M_3**M_3=0$ is the only remaining meson relation.

6th order general matrix relation
\begin{equation}\label{ggm}
\begin{split}
&{M_3}^2+{M_2}^3+{M_1}^6+(M_3M_2M_1+permu.)+(M_3{M_1}^3+permu.)\\
&+({M_2}^2{M_1}^2+permu.)+(M_2{M_1}^4+permu.)=0
\end{split}
\end{equation}

One can show that the meson relation $M_3**M_3=0$ is equivalent to the 6th order general matrix relation, in a similar way as we prove that the 5th order matrix relation is equivalent to $M_2**M_3=0$. The details can be found in Appendix \ref{3333}
The above arguments hold in the general $A_2$-type balanced quivers. As we mentioned before, the trace and matrix relation only rely on shape of mirror quiver, while the meson relation depends on the number in each gauge node. Hence the fact that all the meson relations are equivalent to matrix relations and trace relations reflects that for $A_2$-type quivers the shape exerts strong constraint on the specific data in each node.

\subsection{Why matrix relation terminates?}
In the above proof, we divided the 5th order matrix relation into several parts and then converted each of them into meson relation. These meson relations behave like building blocks of our general matrix relations and trace relations. Now we want to use them to prove the dependence of general matrix relations of different orders, which will help us understand the second property of the general pattern.

\subsubsection{Quiver [3]-(3)-(3)-[3]}

We mentioned that the general matrix relation terminates at order 6, but the 7th order general matrix relation is also correct according to the proposition.

7th order matrix relation is
\begin{equation}
\begin{split}
&{M_1}^7+({M_3}^2M_1+permu.)+(M_3{M_2}^2+permu.)+(M_3M_2{M_1}^2+permu.)+(M_3{M_1}^4\\&+permu.)+({M_2}^3M_1+permu.)+({M_2}^2{M_1}^3+permu.)+(M_2{M_1}^5+permu.)=0
\end{split}
\end{equation}

Every small part in the parentheses corresponds to a meson, for example
\begin{equation}
\begin{split}
{M_3}^2M_1+permu.&=contraction\ of\ M_3**M_3**M_1\\
 M_3{M_2}^2+permu.&=contraction\ of \ M_3**M_2**M_2\\
 M_3M_2{M_1}^2+permu.&=contraction\ of\ M_3**M_2**M_1**M_1\\
 M_3{M_1}^4+permu.&=contraction\ of\ M_3**M_1**M_1**M_1**M_1\\
\ldots
\end{split}
\end{equation}

As the matrices are all $3\times3$, a meson with more than three matrices must be zero. So we only need to consider the first two terms $M_3**M_3**M_1=0$ and $M_3**M_2**M_2=0$. Subsequently, we refer to those mesons that are not trivially equal to zero as relevant ones. The vanishing of these two relevant mesons are actually the result of the previous relations: $M_3**M_3=0$ and $M_3**M_2=0$. Therefore the matrix relation of this order doesn't contain new information, and should not appear in PL.

\subsubsection{Quiver [2]-(3)-(4)-[5]}

Plethystic Logarithm
\begin{equation}
\text{PL}=[11]t+[11]t^2+[11]t^3-t^4-([11]+2)t^5-([11]+2)t^6+\dots
\end{equation}

The three generators are of spin 1, spin 2, spin 3. It has trace relations from 1st order to 6th order, and matrix relations from 5th order to 6th order. These trace and matrix relations are in the standard form according to the propositions we proved. So here we focus on their dependence.  

5th order matrix relation (\ref{gm}) and trace relation (\ref{gtt}) indicate $M_3**M_2=0$.  

6th order trace relation
\begin{equation}
\text{Tr}(\frac{1}{2}{M_3}^2+\frac{1}{3}{M_2}^3+\frac{1}{6}{M_1}^6+M_3M_2M_1+M_3M_1M_2+M_3{M_1}^3+{M_2}^2{M_1}^2+\frac{1}{2}(M_1M_2)^2+M_2{M_1}^4)=0
\end{equation}

The 6th order matrix relation (\ref{ggm}) and trace relation indicate $M_3**M_3=0$.

So the 5th order and 6th order relations are enough to deduce the 7th order matrix relation, since it only contains two sets of relevant terms: ${M_3}^2M_1+permu.$ and $M_3{M_2}^2+permu.$ Therefore this relation is not independent and doesn't show up in the PL. 

\subsubsection{Quiver [1]-(3)-(5)-[7]}

\ \ \ \ Plethystic Logarithm
\begin{equation}
\text{PL}=[11]t+[11]t^2+[11]t^3-t^4-t^5-([11]+2)t^6-t^7+\dots
\end{equation}

It has trace relations from 1st order to 7th order, and matrix relation at 6th order.

7th order trace relation
\begin{equation}
\text{Tr}(\frac{1}{7}{M_1}^7+{M_3}^2M_1+M_3{M_2}^2+(M_3M_2{M_1}^2+\dots)+M_3{M_1}^4+{M_2}^3M_1+({M_2}^2{M_1}^3+\dots)+M_2{M_1}^5)=0
\end{equation}

The 7th order trace relation itself is equivalent to $M_3**M_3**M_1+M_2**M_2**M_3=0$, whose contraction implies the 7th order matrix relation.
	
 The 6th order relations imply $M_3**M_3=0$, which is enough to derive the 8th order relations.

\subsubsection{Quiver (3)-(6)-[9]} 
This is a complete intersection. Its Hilbert Series has a simple form and can be expressed using Plethystic Exponential.
 
Refiend Hilbert Series
\begin{equation}
\begin{split}
\text{HS}=(1-t^4) (1-t^5) (1-t^6) (1-t^7) (1-t^8) (1-t^9)PE[[11]t+[11]t^2+[11]t^3]
\end{split}
\end{equation}

Plethystic Logarithm
\begin{equation}
\text{PL}=[11]t+[11]t^2+[11]t^3-t^4-t^5-t^6-t^7-t^8-t^9
\end{equation}

The PL is finite and it only has trace relations ranging from 1st order to 9th order.  

Similarly, the 7th order matrix relation is a consequence of 7th order trace relation. 
But the relation $M_3**M_3=0$ doesn't hold, due to the absence of 6th order matrix relation. Note that this is also a consequence of the fact that the rightmost node of mirror quiver is 2, while $M_3**M_3=0$ implies this number to be 1.   It follows that the 8th and 9th order trace relations are not trivial. 

\subsubsection{General case}

To count for the third property of the general pattern, note that for all the $A_2$ balanced quivers that are not complete intersections, the rightmost node is 1, while for complete 
intersections the rightmost node is 2. This means $M_k**M_k=0$ should always hold in the former case. It can also be deduced from the 2$k$th order general trace and matrix relations in the non-complete intersection case.  

For $(3k-1)$th order, the only relevant term is ${M_k}^2M_{k-1}+permu.$, and for 3$k$th order, the only relevant term is ${M_k}^3$. Therefore when $M_k**M_k=0$ holds, these two terms are equal to zero as a consequence of this. When $M_k**M_k=0$ doesn't hold, these two relations are not trivial and hence appear in the PL. This explains the third property.

\mbox{}

This method of dividing the matrix relation into small parts could explain the last two properties of general pattern, but here we mention that for a special class of the quivers like [$k$]-($k$)-($k$)-[$k$], we actually don't need the help of these mesons. 

We denote the $n$th order general matrix relation with $k$ generators as $EQM(n,k)$, then the following formula is correct:

\begin{equation}\label{pro}
\begin{split}
&M_k EQM(n-k,k)+M_{k-1} EQM(n-k+1,k)+M_{k-2} EQM(n-k+2,k)\\&+...+M_1 EQM(n-1,k)+EQM(n,k)=0
\end{split}
\end{equation}

This identity can be justified by considering an example. 

For the quiver [$k$]-($k$)-($k$)-[$k$], we already have matrix relations from $(k+1)$th order to 2$k$th order, so we can use this equation to show that the $(2k+1)$th order matrix relation is not independent from the previous ones.

\section{General $A_n$-type quiver: an attempt of complete description}\label{generel}
\subsection{General description}
A-type balanced quivers beyond $A_2$ have the complication that they contain relations that cannot be cast into the form of matrix and trace equations. 

We have mentioned how one could get meson relations and in what representations they transform. In $A_2$ case the generators are all $3\times3$ matrix which means that the second meson (in the form of $M_1**M_2$) is transformed in the same way as matrix relation, and they are equivalent. However, when the size of matrix becomes large, like in $A_3$ case, the second meson relation is no longer equivalent to the matrix relation. Instead, they appear independently. We give a mnemonic to judge which relations are equivalent: as $3=1+2=0+3$, matrix (can be seen as once anti-symmetrized meson) and second meson (twice anti-symmetrized) are `equivalent', and the trace is `equivalent' to third anti-symmetrized meson. In $A_3$ case, $4=2+2=1+3=0+4$, which means second meson is independent and matrix is identified with third meson, trace has the same d.o.f. as the forth meson.

According to this, one could easily predict that a new kind of relation (the relation that transforms in a new representation) will emerge at every $A_n$ with $n$ an odd number. For instance, the independent second meson appears at $A_3$, and the independent third meson appears at $A_5$, etc.

\subsection{One generator case}
One generator quiver, whose smallest gauge node is 1, is the simplest situation. According to Namikawa's theorem \cite{Namikawa2016}, in this case the moduli space is a variety which is the closure of nilpotent orbit of Lie algebra of its isometry group. For example, in the $A_3$ case, the four possible one generator quivers are \{1,2,3\}, \{1,2,2\}, \{1,2,1\}, \{1,1,1\}.  These correspond to the four partitions of 4: \{4\}, \{3,1\}, \{2,2\}, \{2,1,1\} from the maximal orbit to minimal orbit of $\mathfrak{sl}_4$. For the detail, please see \cite{Cabrera:2016vvv}. According to their result, the generator of each quiver can be mapped to a Jordan matrix $X_\lambda$, and therefore the relations can be written down easily.

 Generally, for the quiver whose mirror is ($k_1$)-...-($k_m$)-[$N$], with $m$ gauge nodes ($m\leq k_m$), there are $m$ trace relations: 

\begin{equation}
\text{Tr}M=0,\text{Tr}M^2=0\ldots \text{Tr}M^m=0
\end{equation}

There is a Matrix relation $M^{m+1}=0$ at $(m+1)$th order and a rank relation of order $(k_m+1)$:

\begin{equation}\label{ran1}
\text{rank}{M}\leq k_m
\end{equation}

For integer $m$ , $M^{m+1}=0$ tells us that $\text{rank}(M)\leq [\frac{m}{m+1}N]$\footnote[1]{[$\frac{m}{m+1}N$] denotes integer part of $\frac{m}{m+1}N$.} ($M$ is a $N$ by $N$ matrix). So if $ m<k_m<[\frac{m}{m+1}N]$, a meson relation will appear at $(k_m+1)$th order in the PL, which tells us $\text{rank}(M)\leq k_m$ and puts a stronger constraint on $M$ than the matrix relation does. 

For similar reason, if $m=k_m$, when $k_m<[\frac{m}{m+1}N]$, the $(m+1)$th order relation should be a meson relation instead of a matrix relation. Otherwise, if $k_m\geq [\frac{m}{m+1}N]$, it will be a matrix relation. 

For example, the quiver \{\{1,2,2\},\{0,1,2\}\} with mirror (1)-(2)-[4] has relations $\text{Tr}M=0$, $\text{Tr}(M^2)=0$ and $M^3=0$. The quiver \{\{1,2,2,2\},\{0,1,0,2\}\} with mirror (1)-(2)-[5] has relations Tr$M$=0, $\text{Tr}(M^2)=0$ and $M**M**M=0$. Because this relation means $\text{rank}(M)\leq 2$, while $M^3=0$ only guarantees $\text{rank}(M) \leq [\frac{2}{3}N]=3$.

In 2$(k_{m-1}+1)$th order, there is a relation 
\begin{equation}\label{ran2}
\text{rank}(M^2)\leq k_{m-1}
\end{equation}

This is because $M^2=dudu=dL_1R_1u$, where $R_1L_1$ is a $k_{m-1} \times k_{m-1}$ matrix. One can find the relation $\text{rank}(M^n)\leq k_{m-n+1}$ by the similar reasoning. To see whether they are independent or not, one can write M into the Jordan matrix with all the diagonal elements being zero.  

For example, the quiver with mirror (1)-(4)-[7] has a relation $\text{rank}(M^2)\leq1$, and it is independent from the previous relations.

\subsection{General case}
Generally speaking, the meson relation appears in most of the $A_3$-type quivers, but we still could find a set or more quivers which do not contain meson relations. We start by analyzing the pattern of these quivers.

\subsubsection{Bare pattern: Quiver 242-246}

\begin{table}[htbp]\footnotesize
	\centering
		\begin{tabular}{|c|c|c|l|}
		\hline
		Quiver&Mirror&Relations&Unrefined HS	\\ \hline
242 &  \input{242.tex} & \tabincell{c}{3rd:trace;matrix\\4th:matrix} & \tabincell{l}{$(1+t^2) (1+t+t^2)^2(1+5 t+21 t^2+50 t^3+91 t^4+102 t^5 $ \\ $ +91 t^6+50 t^7+21 t^8+5 t^9+t^{10}))/(1-t)^{16} (1+t)^8$}\\ \hline 
243 &  \input{243.tex} & \tabincell{c}{3rd:trace\\4th:matrix;trace\\5th:matrix} & \tabincell{l}{$(1+t+t^2)^2 (1+2 t+11 t^{2}+21 t^3+50 t^4+71 t^5+92 t^6+71 t^7 $ \\ $+50 t^8+21 t^9+11 t^{10}+2 t^{11}+t^{12})/(1-t)^{18} (1+t)^7$} \\ \hline 
244 &  \input{244.tex} & \tabincell{c}{3rd: trace\\4th:trace\\5th:trace;matrix\\6th:matrix} & \tabincell{l}{$(1+t^2) (1+t+t^2) (1+4 t+14 t^{2}+36 t^{3}+85 t^{4}+159 t^{5} $ \\ $ +260 t^{6}+353 t^7+404 t^8+353 t^9+260 t^{10}+159 t^{11}+85 t^{12}+36 t^{13} $ \\ $+14 t^{14}+4 t^{15}+t^{16})/(1-t)^{20} (1+t)^{10}$} \\ \hline 
245 &  \input{245.tex} & \tabincell{c}{3rd:trace\\4th:trace\\5th:trace\\6th:trace;matrix} & \tabincell{l}{$(1 + t^2) (1 - t + t^2) (1 + t + t^2)^2 (1 + t + t^2 + t^3 + 
   t^4)$\\$ (1 + 4 t^2 + 10 t^4 + 4 t^6 + t^8)/(-1 + t)^{22} (1 + t)^9$}\\ \hline 
246 &  \input{246.tex} & \tabincell{c}{3rd:trace\\4th:trace\\5th:trace\\6th:trace\\7th:trace\\8th:trace}&\tabincell{l}{$(1 - t^3) (1 - t^4) (1 - t^5) (1 - t^6) (1 - t^7)(1 - t^8)/(1 - t)^{15} (1 - t^2)^{15}$} \\ \hline
		\end{tabular}
		\caption{Quiver 242-246}
		\label{tab:242-246}
\end{table}


See Table \ref{tab:242-246}.These mirror quivers all have one gauge node 2 or 3 to the right of flavor node. This means the second meson relation emerges not before 6th order  ($M_2**M_2**M_2=0$). So it will have no chance to interact with the trace and matrix relations before this order. We expect the pattern shares some similarities with the pattern in $A_2$-type quivers. Note that from top to the bottom, the length of left arm of the mirror quiver increases one by one. This could be compared with the $A_2$ quivers from [2]-(2)-(2)-[2] to (2)-(4)-[6] (see Appendix \ref{some}).

Observe that the little difference between this pattern and the $A_2$ pattern is that the matrix relations in this case do not terminate at 4th order. Instead, they shift to higher order gradually and finally terminate at 6th order for (2)-(4)-(5). An obvious explanation is that our discussion of $A_2$-type quiver relies on the condition that the matrix is of $3\times3$, while in the present case with larger matrix size, the pattern need some modification.

Recall that for $m$ generators, $k,\ (k+1),\ (k+2) \ldots\ (k+m-1)$th order general matrix relations together indicate the general matrix relation of $(k+m)$th order (see (\ref{pro})). Therefore, the absence of 5th order relation in first row of the table is easy to understand, since there are two matrix relations present in 3rd and 4th order. 

The 4th order trace relation can be deduced by first multiplying $M_1$ to the 3rd order matrix relation, then take trace:

\begin{equation}\label{mt}
\text{Tr}({M_1}^4+2M_2{M_1}^2)=0 
\end{equation}

Take trace of the 4th order general matrix relation
\begin{equation}\label{4gt}
\text{Tr}({M_1}^4+{M_2}^2+3M_2{M_1}^2)=0
\end{equation}

Combine (\ref{mt}) and (\ref{4gt}), one gets the 4th order general trace relation.

In a complete analogy of the above deduction one can show that the second row and third row of the table also have a similar explanation. Note that they always have two matrix relations, and this is enough to deduce the subsequent matrix relations as well as trace relations. In this sense, these patterns in general $A_n$-type quivers are easier to understand than in the $A_2$ quivers.

\mbox{}

For the last two rows of the table, we do need the information about size of matrix. The fact that the matrices are $4\times4$ allows us to derive the higher order matrix relations by just one matrix equation for quiver 245, and by just trace relations for quiver 246. The proof can be found in Appendix \ref{245}

Note that from the forth row of the table to the last row, the number in rightmost node changes from 2 to 3. This counts the increasing of two trace relations at 7th and 8th order. It is a common feature shared with the $A_2$-type quiver (see the third property of general pattern of $A_2$).

To summarize, the above discussions are quite general for all A-type balanced quivers without the interference of the meson relations. If there are $k$ generators, it will start with $k$ matrix relations and $2k-1$ trace relations. Then the matrix relations shift to higher order gradually with their number fixed, while the number of trace relations increases one by one in this process. Finally, the matrix relation terminates at some order, and the number of trace relations increases by two in the last step. As in this pattern the meson relation's influence is absent,  we refer to it as the bare pattern.

\subsubsection{Modified pattern: Quiver 232-234}

See Table \ref{tab:232-234}. Their mirrors all have a gauge node 1 to the right of flavor node. This gives a second meson relation in 4th order $M_2**M_2=0$ ($(M_2)_{[i}{}^{[j}(M_2)_{k]}{}^{l]}=0$).

\begin{table}[htbp]\footnotesize
	\centering
		\begin{tabular}{|c|c|c|l|}
		\hline
		Quiver&Mirror&Relations&Unrefined HS	\\ \hline
232 &  \input{232.tex} & \tabincell{c}{3rd:trace;matrix\\4th:matrix;$ M_2**M_2=0 $} & \tabincell{l}{$(1+6 t+31 t^{2}+99 t^3+237 t^4+385 t^5+462 t^6$ \\ $ +385 t^7+237 t^8+99 t^9+31 t^{10}+6 t^{11}+t^{12})/(1-t)^{14} (1+t)^5$}\\ \hline 
233 &  \input{233.tex} & \tabincell{c}{3rd:trace\\4th:trace;matrix;\\$M_2**M_2=0$} & \tabincell{l}{ $(1+t+t^2) (1+4 t+19 t^{2}+56 t^3+117 t^4+204 t^5 +214 t^6$ \\ $+204 t^7+117 t^8+56 t^9+19 t^{10}+4 t^{11}+t^{12})/(1-t)^{16} (1+t)^6$}\\ \hline 
234 &  \input{234.tex} & \tabincell{c}{3rd:trace\\4th:trace;$ M_2**M_2=0 $\\5th:trace} & \tabincell{l}{ $(1+t^2)^2 (1+t+t^2) (1+8 t^2+t^4) (1+t+t^2+ t^3+t^4) $ \\ $ /(1-t)^{18} (1+t)^5$}\\ \hline 
		\end{tabular}
		\caption{Quiver 232-234}
		\label{tab:232-234}
\end{table}

Note that the trace relation still increases order by order. However, matrix relation terminates at 4th order in this case. We attribute the difference of this pattern from the previous one to the presence of a second meson relation. To understand how the meson relation interacts with matrix and trace relations and makes them terminate, we focus on the 5th order matrix relation, which has only two  relevant terms: 

\begin{equation}
({M_2}^2M_1+permu.)+(M_2{M_1}^3+permu.)
\end{equation}

The first one is contained in the contraction of 
$M_2**M_2**M_1$, which is zero due to the second meson relation $M_2**M_2=0$. The second one is contained in the contraction of $M_2**M_1**M_1**M_1=0$. There are two ways to deduce  this meson equation, by making use of the 4th order matrix relation or the 5th order trace relation respectively.

Note that the first two rows of this table, quiver 232 and 233, both have 4th order matrix relations. The last row, quiver 234, has 5th order trace relation. Hence for all of them, the 5th order matrix relation can be proved to be dependent on the lower order relations.  The detail of the proof can be found in Appendix \ref{232}.

\textbf{-Another example: Quiver 343-345}

This example contains quivers from (3)-(4)-(3) to (3)-(4)-(5). See Table \ref{tab:343-345}.
\begin{table}[htbp]\footnotesize
	\centering
		\begin{tabular}{|c|c|c|l|}
		\hline
Quiver&Mirror&Relations&Unrefined HS		\\ \hline
343 &  \input{343.tex} & \tabincell{c}{4th:trace;matrix\\5th:trace;matrix;\\$ M_2**M_3=0 $\\6th:matrix;$ M_3**M_3=0 $} &\tabincell{l}{ $(1+t^2) (1+7 t+36 t^{2}+149 t^{3}+517 t^{4}+1545 t^{5}$ \\ $+4038 t^{6}+9249 t^{7}+18614 t^{8}+32948 t^{9}+51320 t^{10}$ \\ $+70372 t^{11}+85039 t^{12}+90576 t^{13}+85039 t^{14}+70372 t^{15}$ \\ $+51320 t^{16}+32948 t^{17}+18614 t^{18}+9249 t^{19}+4038 t^{20}$ \\ $+1545 t^{21}+517 t^{22}+149 t^{23}+36 t^{24}+7 t^{25}+t^{26})/$ \\ $(1-t)^{20} (1+t)^6 (1+t+t^2)^6$}\\ \hline 

344 &  \input{344.tex} & \tabincell{c}{4th:trace\\5th:trace;matrix;\\$ M_2**M_3=0 $\\6th:trace;matrix;\\$M_3**M_3=0$}& \tabincell{l}{ $(1+t^2) (1+8 t+41 t^{2}+169 t^{3}+597 t^{4}+1828 t^{5}$ \\ $+4949 t^{6}+11961 t^{7}+25879 t^{8}+50209 t^{9}+87604 t^{10}$ \\ $+137690 t^{11}+195208 t^{12}+250083 t^{13}+290037 t^{14}$ \\ $+304716 t^{15}+290037 t^{16}+250083 t^{17}+195208 t^{18}$ \\ $+137690 t^{19}+87604 t^{20}+50209 t^{21}+25879 t^{22}+11961 t^{23}$ \\ $+4949 t^{24}+1828 t^{25}+597 t^{26}+169 t^{27}+41 t^{28}+8 t^{29}$ \\ $+t^{30})/(1-t)^{22} (1+t)^9 (1+t+t^2)^6$}\\ \hline 

345 &  \input{345.tex} & \tabincell{c}{4th:trace\\5th:trace;$M_2**M_3=0$\\6th:trace;matrix;\\$M_3**M_3=0$\\7th:trace}& \\ \hline

		\end{tabular}
		\caption{Quiver 343-345}
		\label{tab:343-345}
\end{table}

Note that for these quivers the two second meson relations $M_2**M_3=0$ and $M_3**M_3=0$ are satisfied.
The matrix relation terminates at 6th order. This means we need to prove that the 7th order general matrix relation is not independent from the lower order relations. 

The 7th order matrix relation contains four relevant terms: 

$({M_3}^2M_1+permu.)+(M_3{M_2}^2+premu.)+(M_3M_2M_1^2+permu.)+({M_2}^3M_1+permu.)+\ldots$ 

Again, the first three terms are controlled by $M_3**M_3=0$ and $M_2**M_3=0$ respectively, so we only need to discuss the relation $M_2**M_2**M_2**M_1=0$. Using the 6th order general matrix relation and the two meson relations, one is able to prove that $M_2**M_2**M_2=0$ and $M_3**M_1**M_1**M_1+\frac{3}{2}M_2**M_2**M_1**M_1=0$.  Just $M_2**M_2**M_2=0$ is enough to prove the relation $M_2**M_2**M_2**M_1=0$ (and hence the higher order relations). The proof can be found in Appendix \ref{343}.

In summary, we have seen that independent second meson relations could modify the bare pattern discussed above. We refer to these patterns as the modified ones.

\subsection{Do we have predictability?}

Calculation of Hilbert Series can be very tough for quivers with large size. But fortunately, the method in the above discussion could help us find the relations even when the Hilbert Series is unknown.

We try to give the relations of $A_4$-type quiver with two generators, of which the Hilbert Series is hard to compute. The results are listed in the Appendix \ref{some}.

\section{Balanced D-type quiver}\label{D}

The D-type balanced quiver may also have some interesting patterns. It probably still holds that the smallest node determines the number of generators.  Here we study the simplest examples:
\subsection{2211}
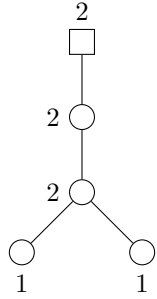
\begin{figure}[h]
	\centering
\begin{tikzpicture}
	\node (g1) at (0,0) [circle,draw,label=left:2] {};
	\node (g2) at (0,1) [circle,draw,label=left:2] {};
	\node (g3) at (-0.8,-0.8) [circle,draw,label=below:1] {};
	\node (g4) at (0.8,-0.8) [circle,draw,label=below:1] {};
	\node (f1) at (0,2)[regular polygon,regular polygon sides=4,draw,label=above:2] {};
	\draw (f1)--(g2)--(g1)--(g3) (g1)--(g4);
\end{tikzpicture}
\caption{quiver 2211}
	\label{fig:fig2211}
\end{figure}
To save space, here we use the symbol \ ${\mu_1}^{n_1}{\mu_2}^{n_2}{\mu_3}^{n_3}{\mu_4}^{n_4}$\ to represent $[n_1,n_2,n_3,n_4]$. The Hibert Series written in this form is called Highest Weight Generating Function (HWG) \cite{Hanany:2014dia}. For the Plethystic Logarithm, It's very important to distinguish between two cases: PL of HS and PL of HWG. They are different because of the different multiplication rules. In the former case the multiplication of two representation, for example $\mu_2 \times\mu_2$, is ${\mu_2}^2+\mu_1\mu_3\mu_4+{\mu_1}^2+{\mu_3}^2+{\mu_4}^2+\mu_2+1$, while in the latter case this product is ${\mu_2}^2$. To avoid confusion, in the following we denote the fugacity of HWG and PL of HWG by $\mu$, and denote the fugacity of HS and PL of HS by $\nu$.

The PL of HS is
\begin{equation}
\nu_1 t+\left(-{\nu_3}^2-{\nu_4}^2-1\right) t^2+\left({\nu_3}^2+\nu_1 \nu_4 \nu_3+{\nu_4}^2\right) t^3+\dots
\end{equation}

So there is only one generator of spin 1,  and relations of 2nd order in representation $1+[0020]+[0002]$.

The mirror of this theory is nilpotent orbit $[D_4]-(C_1)-(B_0)$. In the following, i,j=1,2,...8 is in the SO(8) group, $\alpha,\beta=1,2$ in the SP(1) group,  $\Omega$ is the invariant tensor of Symplectic group. $A_{i\alpha}$ is the half-hypers between O(8) and SP(1), $B_{\alpha}$ is the half-hypers between SP(1) and O(1).

Generator:
\begin{equation}
M=A\Omega A^T,\ 8\times8\ \text{matrix}
\end{equation}

F-term equations:
\begin{equation}
A^TA=BB^T, 
\end{equation}
and
\begin{equation}	
 B^T\Omega B=0
\end{equation}

We denote the generator as $M$, which is a 8 by 8 matrix with rank 2. So the Pfaffian of 4 by 4 minor vanishes: 
\begin{equation}
\epsilon _{j_1 j_2 j_3 j_4}^{i_1 i_2 i_3 i_4} M^{j_1 j_2} M^{j_3 j_4}=0
\end{equation}

This relation is in [0020]+[0002]. In addition to this relation, there is a trace relation $\text{Tr}(M^2)$ is this order.

\subsection{2322}
\begin{figure}[h]
	\centering
\begin{tikzpicture}
	\node (g1) at (0,0) [circle,draw,label=left:3] {};
	\node (g2) at (0,1) [circle,draw,label=left:2] {};
	\node (g3) at (-0.8,-0.8) [circle,draw,label=below:2] {};
	\node (g4) at (0.8,-0.8) [circle,draw,label=below:2] {};
	\node (f1) at (0,2)[regular polygon,regular polygon sides=4,draw,label=left:1] {};
	\node (f2) at (-1.6,-1.6)[regular polygon,regular polygon sides=4,draw,label=below:1] {};
	\node (f3) at (1.6,-1.6)[regular polygon,regular polygon sides=4,draw,label=below:1] {};
	\draw (f1)--(g2)--(g1)--(g3)--(f2) (g1)--(g4)--(f3);
\end{tikzpicture}
\caption{quiver 2322}
	\label{fig:fig2322}
\end{figure}
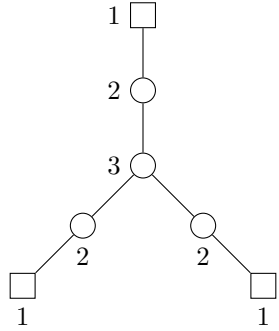

In this example, the Highest Weight Generating function has a nice form:
\begin{equation}\label{hwg}
\begin{split}
\text{HWG}=&\frac{1-{\mu_1}^2 {\mu_3}^2 {\mu_4}^2 t^6}{\left(1-{\mu_2} t\right) \left(1-{\mu_1}^2 t^2\right) \left(1-{\mu_2} t^2\right) \left(1-{\mu_3}^2 t^2\right) \left(1-{\mu_1} {\mu_3} {\mu_4} t^3\right){}^2 \left(1-{\mu_4}^2 t^2\right)}\\
\\
=&1+{\mu_2} t+\left({{\mu_1}}^2+{\mu_2}^2+{\mu_3}^2+{\mu_4}^2+{\mu_2}\right) t^2+({\mu_2}^3+{\mu_2}^2+{\mu_1}^2 {\mu_2}+{\mu_3}^2 {\mu_2}\\&+{\mu_4}^2 {\mu_2}+2 {\mu_1} {\mu_3} {\mu_4}) t^3+({\mu_2}^4+{\mu_1}^2{\mu_2}^2+{\mu_2}^2{\mu_3}^2+{\mu_2}^2{\mu_4}^2+{\mu_2}^3+2{\mu_1}{\mu_2}{\mu_3}{\mu_4}\\&+{\mu_1}^4+{\mu_3}^4+{\mu_4}^4+{\mu_1}^2{\mu_3}^2+{\mu_1}^2{\mu_4}^2+{\mu_3}^2{\mu_4}^2+{\mu_2}{\mu_1}^2+{\mu_2}{\mu_3}^2+{\mu_2}{\mu_4}^2+{\mu_2}^2)t^4+\dots 
\end{split}
\end{equation}

The PL of HS is
\begin{equation}\label{pl}
{\nu_2} t+\left({\nu_2}-1\right) t^2+\left(-{\nu_1}^2-{\nu_3}^2-{\nu_4}^2-3 {\nu_2}-1\right) t^3+\left({\nu_1}^2-{\nu_2}^2+{\nu_3}^2+{\nu_4}^2+3 {\nu_2}\right) t^4+\dots
\end{equation}

This Hilbert Series coincide with the HS of Higgs branch of the theory $[SO(8)]-(C_2)-(SO(2))$, which can be calculated by first writing down the Hilbert Series without gauge invariant constraint and then doing Molien-Weyl integration to project the operators to gauge invariant terms \cite{Hanany:2016gbz}:
\begin{equation} 
\label{BCD}
\begin{aligned}
\text{HS}_{Higgs}^{[SO(8)]-C_2-SO(2)}&= {\oint\limits_{{SO(2), C_2}} {d\mu } \frac{{PE\left[[vector]_{SO(8)}\otimes[fund]_{C_2}t+ {{{\left[ {fund} \right]}_{{C_2}}} \otimes {{\left[ {vec} \right]}_{{SO(2)}}}t} \right]}}{{PE\left[ [adjoint]_{SO(2)}t^2 +[adjoint]_{C_2}t^2\right]}}}\\
 \end{aligned}
\end{equation}

The unrefiend calculation shows that it is the same as the 2322 Coulomb branch HS. It indicates the two moduli space are the same.  Denoting the half-hypers between SO(8) and $C_2$ as 
\begin{equation}
A_{i\alpha}, \ \ \ i\ from\ 1\ to\ 8\ in\ SO(8),\ \alpha\ from\ 1\ to\ 4\ in\ C_2
\end{equation}
and the half-hypers between $C_2$ and SO(2) as 
\begin{equation}
B_{\alpha r}, \ \ \ \alpha\ from\ 1\ to\ 4\ in\ C_2,\ r\ from\ 1\ to\ 2\ in\ SO(2)
\end{equation}

These two operators correspond to the numerator of integrand of Eq.(\ref{BCD}). They satisfy the F-term equations  
\begin{equation}\label{ftm}
\begin{split}
&A^TA=BB^T\\
&B^T\Omega B=0
\end{split}
\end{equation}

These two equations correspond to the denominator of the integrand of Eq.(\ref{BCD}). From the quiver fields we can construct two independent generators $M_1$ and $M_2$ with spin one and spin two respectively
\begin{equation}
\begin{split}
&M_1=A\Omega A^T\\
&M_2=A\Omega B \epsilon B^T\Omega A^T
\end{split}
\end{equation}

Note that the second generator changes sign under $Z_2$ action which interchanges $B_{\alpha 1}$ with $B_{\alpha 2}$ and thus is not invariant under the full group of O(2). But it is an invariant of the gauge group SO(2).   

Most of the relations can be found by merely looking at the two functions (\ref{hwg}) and (\ref{pl}). The prescription is that when an operator is constructed transforming under certain representation, and if the HS (HWG) doesn't contain this term, one can conclude that it must vanish due to some relations. By this method, we can find all the following relations.  

\textbf{Relations}

2nd order
\begin{equation}\label{2rr}
\text{Tr}({M_1}^2)=0
\end{equation}

3rd order
\begin{equation}\label{3rr}
\begin{split}
&[2000]+[0000]:\  M_1M_2+M_2M_1=0\\
&[0020]+[0002]:\ \epsilon _{i_1 i_2 i_3 i_4 j_1 j_2 j_3 j_4} (M_1)^{j_1 j_2} (M_2)^{j_3 j_4}=0\\
&3[0100]:\qquad \quad \ M_1M_2-M_2M_1=0,  \  {M_1}^3=0,    \ \epsilon _{i_1 i_2 j_1 j_2 j_3 j_4 j_5 j_6} (M_1)^{j_1 j_2} (M_1)^{j_3 j_4} (M_1)^{j_5 j_6}=0
\end{split}
\end{equation}

In 4th order, there are two nontrivial terms in chiral ring transforming under [0200] and one combination of them is relation. One [0200] can be constructed purely from $M_1$:
\begin{equation}\label{re1}
(M_1^2)_{['i_1[i_3}(M_1^2)_{i_2]'i_4]}
\end{equation}

The index structure implies that it transforms under 
\begin{equation}
sym^2[0100]=[2000]+[0200]+[0020]+[0002]
\end{equation}

However, when two indices are contracted, one gets the combination of ${M_1}^4$ and $\text{Tr}({M_1}^2){M_1}^2$. All of them equal zero as a consequence of the lower order relations. Antisymmetrizing the four indices also gives zero since ${M_1}^2$ is symmetric matrix. Therefore, the expression (\ref{re1}) actually transforms under representation [0200].

Another term can be constructed from $M_2$:
\begin{equation}\label{re2}
\begin{split}
&(M_2)_{i_1i_2}(M_2)_{i_3i_4}-\epsilon_{i_1i_2i_3i_4}^{i_5i_6i_7i_8}(M_2)_{i_5i_6}(M_2)_{i_7i_8}+
\frac{1}{6}({M_2}^2)_{i_1i_3}\delta_{i_2i_4}-\frac{1}{6}({M_2}^2)_{i_1i_4}\delta_{i_2i_3}-\frac{1}{6}({M_2}^2)_{i_2i_3}\delta_{i_1i_4}\\&+\frac{1}{6}({M_2}^2)_{i_2i_4}\delta_{i_1i_3}-\text{Tr}({M_2}^2)(\frac{1}{42}\delta_{i_1i_3}\delta_{i_2i_4}-\frac{1}{42}\delta_{i_1i_4}\delta_{i_2i_3})
\end{split}
\end{equation}

This also transforms under [0200]. Note that in the 4th order of HWG (\ref{hwg}), no term transforms under [2000] and [0020]+[0002]. This implies that $(M_2)^2=0$ and $\epsilon_{i_1i_2i_3i_4}^{i_5i_6i_7i_8}(M_2)_{i_5i_6}(M_2)_{i_7i_8}=0$. As a consequence, we can simplify the expression (\ref{re2}) to 
\begin{equation}\label{re3}
(M_2)_{i_1i_2}(M_2)_{i_3i_4}
\end{equation}

The [0200] relation appearing in 4th order of PL should be a combination of these two terms (\ref{re1}) and (\ref{re3}), which is parimetrized by a constant $a$
\begin{equation}\label{relf}
(M_1^2)_{['i_1[i_3}(M_1^2)_{i_2]'i_4]}+a(M_2)_{i_1i_2}(M_2)_{i_3i_4}=0
\end{equation}

By analyzing the Higgs branch of theory [SO(8)]-$C_2$-(SO(2)), $a$ is determined to be $-\frac{1}{2}$. The proof of these relations can be found in Appendix \ref{higgs}.
 
Interestingly the moduli space of this theory coincides with the moduli space of nilpotent orbit $\{3,2^2,1\}$\footnote{this is the partition of the theory as used in \cite{Hanany:2016gbz}} when the generator $M_2$ is set to zero.  

The HWG of theory $\{3,2^2,1\}$ can be found in \cite{Hanany:2016gbz}:
\begin{equation}
\text{HWG}=\frac{1 + \mu_1\mu_3\mu_4t^3}{(1 - \mu_2t)(1 - {\mu_1}^2t^2)(1-{\mu_3}^2t^2) (1 - {\mu_4}^2t^2)}
\end{equation}

Denote the only generator as $M$. The relations are:

2nd order: 
\begin{equation}
\text{Tr}{M^2}=0
\end{equation}

3rd order:
\begin{equation}
\begin{split}
&M^3=0\\
\epsilon _{i_1 i_2 j_1 j_2 j_3 j_4 j_5 j_6} M&^{j_1 j_2} M^{j_3 j_4} M^{j_5 j_6}=0
\end{split}
\end{equation}

4th order:
\begin{equation}\label{rank}
\text{rank}(M^2)\leq1
\end{equation}

So if we set $M_2=0$ in quiver 2322, the relations (\ref{2rr}), (\ref{3rr}), (\ref{relf}) of $M_1$ match exactly with the relations of $M$. On the other hand, considering the relations composed only of $M_2$: 
\begin{equation}
\begin{split}
(M_2)^2&=0\\
\epsilon_{i_1i_2i_3i_4}^{i_5i_6i_7i_8}(M_2)_{i_5i_6}(M_2)_{i_7i_8}&=0
\end{split}
\end{equation} 
which implies that the moduli space generated by $M_2$ alone coincides with the moduli space of one SO(8)-instanton. Therefore the moduli space of 2322 can be seen as the two matrices generalization of nilpotent orbit, with the fourth order rank relation of $M_1$ (\ref{rank}) being corrected by $M_2$ (see (\ref{relf})).   

\section*{Acknowledgments}
G. C., Y. L. and Y. Z. are very grateful to Santiago Cabrera, Rudolph Kalveks, Giulia Ferlito and Zhenghao Zhong for helpful conversations, and to the USTC/Imperial Undergraduate Research Program that made possible this collaboration. A. H. would like to thank the Indian Institute of Science Education and Research Pune, and Tata Institute of Fundamental Research at Mumbai for their kind hospitality during the last stages of this project. A. H. is supported by STFC Consolidated Grant ST/J0003533/1, and EPSRC Programme Grant EP/K034456/1.

\appendix
 
\section{Details of calculations}

\subsection{Quiver [2]-(2)-(2)-[2]}\label{2222}

\textbf{3rd order}

First, we prove that the matrix and trace relation of this order imply second meson and determinant relation.

$\text{Tr}(\frac{1}{3} {M_1}^3+M_1 M_2)=0$ and $\text{Tr}({M_1}^3+2M_1 M_2)=0$ (trace of (\ref{any})) indicate $\text{Tr}({M_1}^3)=0$. We also know that $\text{Tr}(M_1)=0$, so $\text{Det}(M_1)=0$. Because $x_1+x_2+x_3=0$ and ${x_1}^3+{x_2}^3+{x_3}^3=0$ imply $x_1x_2x_3$=0.

Using Cayley-Hamilton formula\footnote{Since Tr($M_1$)=0, $p_2(\lambda)=\lambda_1\lambda_2+\lambda_1\lambda_3+\lambda_2\lambda_3=-\frac{1}{2}\text{Tr}({M_1}^2)$, $\text{Det}(M_1)=\frac{1}{3}\text{Tr}({M_1}^3)$}  
\begin{equation}
{M_1}^3-\frac{1}{2} \text{Tr}({M_1}^2)M_1=0
\end{equation}

and the 3rd order general matrix relation 
\begin{equation}
{M_1}^3+M_1 M_2+M_2 M_1=0
\end{equation}

one gets 
\begin{equation}
-\text{Tr}(M_2) M_1+M_1 M_2+M_2 M_1=0
\end{equation}
 
This is the just the contraction of meson relation (\ref{me}) we promise to derive. As every step is reversible, the matrix relation of 3rd order is equivalent to the second meson relation $(M_1)_{[i}{}^{[j}(M_2)_{k]}{}^{l]}=0$. 

\textbf{4th order}

Now, we give a proof that the 4h order matrix relation is equivalent to the meson relation $M_2**M_2=0$.

Multiply the 3rd order general matrix relation by $M_1$ and take trace of it
\begin{equation}
\text{Tr}({M_1}^4+2{M_1}^2M_2)=0
\end{equation}

This is equivalent to $M_1**M_1**M_2=0$.

As we can contract it twice 
\begin{equation}
-\text{Tr}({M_1}^2)\text{Tr}(M_2)+2\text{Tr}( {M_1}^2M_2)=0
\end{equation}

The equivalence follows from Cayley-Hamilton formula 
\begin{equation}
{M_1}^4-\frac{1}{2}\text{Tr}({M_1}^2){M_1}^2=0
\end{equation}

Contract  $M_1**M_1**M_2=0$ once, we get a matrix relation
\begin{equation}
-2 \text{tr}({M_1}^2)M_2-4 \text{tr}(M_1 M_2)M_1+4 (M_2 {M_1}^2+ {M_1}^2M_2+M_1 M_2 M_1)-4 {M_1}^2 
\text{Tr}M_2=0
\end{equation}

Combine this with the matrix relation (\ref{4lgm}), one can derive 
\begin{equation}
{M_2}^2-\text{Tr}(M_2)M_2=0
\end{equation}

This is just the contraction of $M_2**M_2=0$.

\subsection{Quiver [3]-(3)-(3)-[3]}\label{3333}
The mirror is
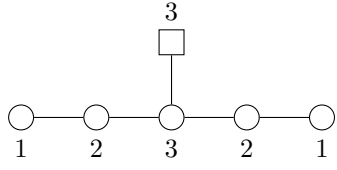
\begin{figure}[h]
	\centering
		\begin{tikzpicture}
	\node (g1) at (-2,0) [circle,draw,label=below:1] {};
	\node (g2) at (-1,0) [circle,draw,label=below:2] {};
	\node (g3) at (0,0) [circle,draw,label=below:3] {};
	\node (g4) at (1,0)[circle,draw,label=below:2] {};
	\node (g5) at (2,0)[circle,draw,label=below:1] {};
	\node (f1) at (0,1)[regular polygon,regular polygon sides=4,draw,label=above:3] {};
	\draw (g1)--(g2) --(g3)--(g4)--(g5)
				(f1)--(g3);
\end{tikzpicture}
	\label{fig:fig6}
	\caption{Mirror of quiver (3)-(3)}
\end{figure}

First, We prove the relation $M_1**M_1**M_2=0$,

\textit{Proof}:
$$(M_1)_i{}^j=d_i{}^\alpha u_\beta {}^j \delta_\alpha {}^\beta,i,j=1,2,3;\alpha,\beta=1,2,3;$$
$$(M_2)_k{}^l=d_k{}^\gamma u_\kappa{}^l Q_\gamma{}^\kappa,\text{where}\ Q_\gamma{}^\kappa=(r_1l_1)_\gamma{}^\kappa=(r_1)_\gamma{}^1(l_1)_1{}^\kappa$$

\begin{equation}
\begin{split}
(M_1)_{[i}{}^{[j}(M_1)_{m}{}^{n}(M_2)_{k]}{}^{l]} &= d_{[i}{}^\alpha d_{m}{}^\mu d_{k]}{}^\gamma u_\beta {}^{[j}u_\nu {}^{n} u_\kappa{}^{l]} \delta_\alpha {}^\beta \delta_\mu {}^\nu Q_\gamma{}^\kappa\\
&=d_{i}{}^{[\alpha} d_{m}{}^\mu d_{k}{}^{\gamma]} u_{[\beta} {}^{j}u_{\nu} {}^{n} u_{\kappa]}{}^{l} \delta_\alpha {}^\beta \delta_\mu {}^\nu Q_\gamma{}^\kappa\\
&=d_{i}{}^{\alpha} d_{m}{}^\mu d_{k}{}^{\gamma} u_{\beta} {}^{j}u_{\nu} {}^{n} u_{\kappa}{}^{l} \delta_{[\alpha} {}^{[\beta} \delta_\mu {}^\nu Q_{\gamma]}{}^{\kappa]}\\
\end{split}
\end{equation}

and 
\begin{equation}
\begin{split}
\delta_{[\alpha} {}^{[\beta} \delta_\mu {}^\nu Q_{\gamma]}{}^{\kappa]}&=\epsilon_{\beta\nu\kappa}\epsilon^{\alpha\mu\gamma}\delta_\alpha {}^\beta\delta_\mu {}^\nu Q_\gamma{}^\kappa\\
&=\epsilon_{\alpha\mu\kappa}\epsilon^{\alpha\mu\gamma} Q_\gamma{}^\kappa\\
&=2\delta_\kappa{}^\gamma Q_\gamma{}^\kappa\\
&=2\text{Tr}(Q)
\end{split}
\end{equation}

Using F-term, we have $\text{Tr}(Q)=0$.

This relation can be derived from general matrix and trace relation (\ref{4gm}) and (\ref{r}). To see this, take trace of Eq.(\ref{4gm}), and combine it with Eq.(\ref{r}) to eliminate $\text{Tr}(M_1M_3)$ and $\text{Tr}{(M_2^2)}$
\begin{equation}
\text{Tr}(\frac{1}{4}{M_1}^4+\frac{1}{2}{M_1}^2M_2)=0
\end{equation}

Then using Cayley-Hamilton theorem, we arrive at Eq.(\ref{ct}) again. This is equivalent to $M_1**M_1**M_2=0$.

Contract it once
\begin{equation}
-2 \text{Tr}({M_1}^2)M_2-4 \text{Tr}(M_1 M_2)M_1+4 (M_2 {M_1}^2+ {M_1}^2M_2+M_1 M_2 M_1)-4 {M_1}^2 
\text{Tr}M_2=0
\end{equation}
and solve from it that
\begin{equation}
M_2 {M_1}^2+ {M_1}^2M_2+M_1 M_2 M_1={M_1}^2 \text{Tr}M_2+\text{Tr}(M_1 M_2)M_1+\frac{1}{2}\text{Tr}({M_1}^2)M_2
\end{equation}

Substitute this to the general matrix relation of 4th order (\ref{4gm}), one gets:
\begin{equation}
{M_1}^4+{M_2}^2+(M_1M_3+M_3M_1)+{M_1}^2TrM_2+\text{Tr}(M_1 M_2)M_1+\frac{1}{2}\text{Tr}({M_1}^2)M_2=0
\end{equation}

Then use the Cayley-Hamilton formula:
\begin{equation}
{M_1}^4-\frac{1}{2}\text{Tr}({M_1}^2){M_1}^2-\frac{1}{3}\text{Tr}({M_1}^3)M_1=0
\end{equation}
we get:
\begin{equation}
{M_2}^2+(M_1M_3+M_3M_1)-\text{Tr}(M_2)M_2-\text{Tr}(M_3)M_1=0
\end{equation}

This is just the contraction of $\frac{1}{2}M_2**M_2+M_1**M_3=0 $.

In the last step, we have used 
\begin{equation}
\text{Tr}(\frac{1}{2}{M_1}^2+M_2)=0 
\end{equation}
and
\begin{equation}
\text{Tr}(M_3+M_1M_2+\frac{1}{3}{M_1}^3)=0
\end{equation}

\mbox{}

\textbf{5th order}

The contraction of $M_1**M_1**M_1**M_2=0$ is
\begin{equation}\label{longe}
\begin{split}
&-({M_1}^3M_2+permu.)+\frac{1}{3}\text{Tr}({M_1}^3)M_2+\text{Tr}({M_1}^2M_2)M_1+\text{Tr}(M_1M_2){M_1}^2\\&+\frac{1}{2}(M_1M_2+M_2M_1)\text{Tr}({M_1}^2)
-\frac{1}{2}\text{Tr}({M_1}^2)\text{Tr}(M_2)M_1+(\text{Tr}M_2){M_1}^3=0
\end{split}
\end{equation}

The contraction of $M_2**M_2**M_1=0$ is
\begin{equation}\label{nl}
\begin{split}
&-4\text{Tr}(M_1M_2)M_2-2\text{Tr}({M_2}^2)M_1+4(M_1{M_2}^2+M_2M_1M_2+{M_2}^2M_1)\\&+2(\text{Tr}(M_2))^2M_1-4(M_1M_2+M_2M_1)\text{Tr}(M_2)=0
\end{split}
\end{equation}

The contraction of $M_2**M_3=0$ is
\begin{equation}\label{st}
M_2M_3+M_3M_2-\text{Tr}(M_2)M_3-\text{Tr}(M_3)M_2=0
\end{equation}

The contraction of $M_3**M_1**M_1=0$ is
\begin{equation}\label{nst}
-2 \text{Tr}({M_1}^2)M_3-4 \text{Tr}(M_1 M_3)M_1+4 (M_3 {M_1}^2+ {M_1}^2M_3+M_1 M_3 M_1)-4 {M_1}^2 
\text{Tr}M_3=0
\end{equation}

Add them together, one gets:
\begin{equation}\label{md}
\begin{split}
&({M_1}^3M_2+permu.)+\frac{1}{3}\text{Tr}({M_1}^3){M_1}^2+\frac{1}{2}\text{Tr}({M_1}^2)\text{Tr}(M_2)M_1-(\text{Tr}M_2){M_1}^3\\&+(M_1{M_2}^2+M_2M_1M_2+{M_2}^2M_1)+\frac{1}{2}(\text{Tr}(M_2))^2M_1+\frac{1}{4}\text{Tr}({M_1}^4)M_1+M_2M_3\\&+M_3M_2+(M_3 {M_1}^2+ {M_1}^2M_3+M_1 M_3 M_1)=0 
\end{split}
\end{equation}

Note that from Cayley-Hamilton formula:
\begin{equation}
{M_1}^4-\frac{1}{2}\text{Tr}({M_1}^2){M_1}^2-\frac{1}{3}\text{Tr}({M_1}^3)M_1=0
\end{equation}

Take trace, we get
\begin{equation}
\text{Tr}({M_1}^4)=\frac{1}{2}(\text{Tr}({M_1}^2))^2
\end{equation}

Substitute to Eq.(\ref{md}).

And observe that
\begin{equation}
\begin{split}
&\frac{1}{3}\text{Tr}({M_1}^3){M_1}^2+\frac{1}{2}\text{Tr}({M_1}^2)\text{Tr}(M_2)M_1-(\text{Tr}M_2){M_1}^3\\&+\frac{1}{2}(\text{Tr}(M_2))^2M_1+\frac{1}{8}(\text{Tr}({M_1}^2))^2M_1\\
=&\frac{1}{3}\text{Tr}({M_1}^3){M_1}^2-(\text{Tr}M_2){M_1}^3\\
=&{M_1}^5
\end{split}
\end{equation}

Therefore, Eq.(\ref{md}) is just the 5th order general matrix relation.

\mbox{}

Now we want to use some trace equations to derive $M_2**M_2**M_1=0$ and $M_3**M_1**M_1=0$.

5th order general trace relation is
\begin{equation}\label{yn}
\text{Tr}(\frac{1}{5}{M_1}^5+M_2{M_1}^3+{M_2}^2M_1+M_3M_2+M_3{M_1}^2)=0
\end{equation}

Also make use of the 5th order general matrix relation and 4th order general matrix relation (multiplied by $M_1$). Taking trace of them
\begin{equation}\label{en}
\text{Tr}({M_1}^5+4M_2{M_1}^3+3{M_2}^2M_1+2M_3M_2+3M_3{M_1}^2)=0
\end{equation} 
\begin{equation}\label{sn}
\text{Tr}({M_1}^5+3M_2{M_1}^3+{M_2}^2M_1+2M_3{M_1}^2)=0
\end{equation}

With these three equations, we can get rid of the terms $\text{Tr}(M_3{M_1}^2)$ and $\text{Tr}(M_3M_2)$ to obtain
\begin{equation}
\frac{1}{10}\text{Tr}({M_1}^5)+\frac{1}{2}\text{Tr}(M_2{M_1}^3)+\frac{1}{2}\text{Tr}({M_2}^2M_1)=0
\end{equation}

Then take trace of Eq.(\ref{longe})
\begin{equation}
\text{Tr}(M_2{M_1}^3)=\frac{1}{3}\text{Tr}({M_1}^3)\text{Tr}(M_2)+\frac{1}{2}\text{Tr}(M_1M_2)\text{Tr}({M_1}^2)
\end{equation}

Substitute it in last equation, one gets
\begin{equation}
-\text{Tr}(M_1M_2)\text{Tr}(M_2)+\text{Tr}({M_2}^2M_1)=0
\end{equation}

This is just the contraction of $M_2**M_2**M_1=0$.

In the last step, we have used the relation $\text{Tr}(\frac{1}{2}{M_1}^2+M_2)=0$ 
and Cayley-Hamilton theorem
\begin{equation}
\text{Tr}[{M_1}^2({M_1}^3-\frac{1}{2}\text{Tr}({M_1}^2)M_1-\text{Det}M_1)]=0
\end{equation}

It's a similar procedure to get rid of the terms $\text{Tr}({M_2}^2M_1)$ and $\text{Tr}(M_3M_2)$ from Eq.(\ref{yn})(\ref{en}) and (\ref{sn}). Then one can derive that contraction of $M_1**M_1**M_3=0$.

\textbf{6th order}

We prove that $M_3**M_3$ can be derived from the 6th order general matrix relation
\begin{equation}\label{6gmm}
\begin{split}
&{M_3}^2+{M_2}^3+{M_1}^6+(M_3M_2M_1+permu.)+(M_3{M_1}^3+permu.)\\
&+({M_2}^2{M_1}^2+permu.)+(M_2{M_1}^4+permu.)=0
\end{split}
\end{equation}
	
We already know that $M_2**M_2**M_2=0$ and $M_1**M_2**M_3=0$ as consequences of lower order relations. 

$M_3**M_1**M_1**M_1$,    $M_2**M_2**M_1**M_1=0$ and $M_2**M_1**M_1**M_1**M_1=0$ trivially hold since the matrices are of $3\times3$.

Similar as in 5th order, the contractions of these terms are the constituents of (\ref{6gmm}). Hence, contraction of $M_3**M_3=0$ can be deduced by subtracting (\ref{6gmm})  by these contractions.

\subsection{Quiver 245-246}\label{245}
Here, we mainly consider the quivers 245 and 246, as their relations are affected by the size of the matrix. 

Consider 245 first. 
6th order matrix relation
\begin{equation}\label{6gm}
{M_1}^6+{M_2}^3+({M_2}^2{M_1}^2+permu.)+\ldots=0
\end{equation}

Take trace of it
\begin{equation}\label{t6m}
\text{Tr}({M_1}^6+{M_2}^3+2(M_2M_1)^2+4{M_2}^2{M_1}^2+\ldots)=0
\end{equation}

General trace relation of this order is
\begin{equation}
\text{Tr}(\frac{1}{6}{M_1}^6+\frac{1}{3}{M_2}^3+\frac{1}{2}(M_2M_1)^2+{M_2}^2{M_1}^2+\ldots)=0
\end{equation}

Combine these two equations to eliminate the term $\text{Tr}({M_2}^3)$, one gets
\begin{equation}
\text{Tr}(\frac{1}{2}{M_1}^6+\frac{1}{2}(M_2M_1)^2+{M_2}^2{M_1}^2+\ldots)=0
\end{equation}

Note that in this equation, only $\text{Tr}(\frac{1}{2}(M_2M_1)^2+{M_2}^2{M_1}^2)$ is relevant.

It indicates  $M_2**M_2**M_1**M_1=0$.

Contract it for three times, we get a matrix relation that contains ${M_2}^2{M_1}^2+permu.$

Substitute it in Eq.(\ref{6gm}), and note that at 6th order only ${M_2}^2{M_1}^2+permu.+{M_2}^3$ are relevant. Therefore $M_2**M_2**M_2=0$. With this relation, it is easy to show the 7th order relations are trivial.
 
For quiver 246, we do not have the 6th order matrix relation. Hence the above argument doesn't hold in this case and we cannot deduce the relation $M_2**M_2**M_2=0$. In fact, this relation cannot be correct since the rightmost node of mirror quiver is 3. 

As a result, the 7th and 8th trace relations 
\begin{equation}
\begin{split}
\text{Tr}({M_2}^3M_1+\ldots)=0\\
\text{Tr}({M_2}^4+\ldots)=0
\end{split}
\end{equation}
are not trivial. Note that we only write the relevant terms, and there is just one of them in both 7th and 8th order. Therefore, the matrix relations of these orders can be derived easily from these trace relations.

\subsection{Quiver 232-234}\label{232}

We want to prove that the 5th order matrix relation can be deduced by the lower order relations.

For this purpose, we only need to deduce the relation $M_2**M_1**M_1**M_1=0$ as discussed in article.

4th order relations
\begin{equation}\label{4g}
\begin{split}
{M_1}^4+{M_2}^2+(M_2{M_1}^2+M_1M_2M_1+{M_1}^2M_2)=0\\
\text{Tr}(\frac{1}{4}{M_1}^4+\frac{1}{2}{M_2}^2+M_2{M_1}^2)=0
\end{split}
\end{equation}

Take trace of the first	equation. Combine it with the second to eliminate $\text{Tr}({M_1}^2M_2)$, one gets
\begin{equation}
\text{Tr}(-\frac{1}{2}{M_2}^2+\frac{1}{4}{M_1}^4)=0
\end{equation}

Combining it with $M_2**M_2=0$, we get $M_1**M_1**M_1**M_1=0$.

Contract this relation to a matrix relation. After substituting it back to the matrix equation in (\ref{4g}), one finds $M_2**M_1**M_1=0$.

For quiver 232, we don't have 4th order trace relation. However, one can multiply the 3rd order matrix relation by $M_1$ and take trace of it to get another trace equation.  Then the deduction is similar as before. 	   

For quiver 234, the 4th order matrix relation is absent. But the 5th order trace relation can be utilized.

5th order trace relation is 
\begin{equation}
\text{Tr}({M_2}^2M_1+M_2{M_1}^3+\ldots)=0
\end{equation}

Note that the first term $\text{Tr}({M_2}^2M_1)$ is trivial since $M_2**M_2=0$. Therefore, only second term is relevant.  This trace relation is equivalent to the relation $M_2**M_1**M_1**M_1=0$.

\subsection{Quiver 343-345}\label{343}
 We want to prove that the two meson relations
\begin{equation}
\begin{split}
M_2**M_3=0\\
M_3**M_3=0
\end{split}
\end{equation}

make the matrix relations terminate at order 6.

As discussed in article, we only need to analyze the 7th order relation 
\begin{equation}
M_2**M_2**M_2**M_1=0
\end{equation}

Note that all of these quivers have the 6th order matrix relation 
\begin{equation}\label{6gm2}
\begin{split}
&{M_3}^2+{M_2}^3+{M_1}^6+(M_3M_2M_1+permu.)+(M_3{M_1}^3+permu.)\\
&+({M_2}^2{M_1}^2+permu.)+(M_2{M_1}^4+permu.)=0
\end{split}
\end{equation}

There is a trace equation, either from  5th order matrix relation multiplied by $M_1$, or from the 6th order general trace relation.

For example, the 6th order trace relation is 
\begin{equation}\label{6ttr}
\begin{split}
&\text{Tr}(\frac{1}{2}{M_3}^2+\frac{1}{3}{M_2}^3+\frac{1}{6}{M_1}^6+M_3M_2M_1+M_3M_1M_2+M_3{M_1}^3\\
&+{M_2}^2{M_1}^2+\frac{1}{2}(M_2M_1)^2+M_2{M_1}^4)=0
\end{split}
\end{equation}

Take trace of (\ref{6gm2}). Eliminate $\text{Tr}({M_2}^3)$ using (\ref{6ttr}), the resulting equation is 
\begin{equation}
\text{Tr}(M_3{M_1}^3+{M_2}^2{M_1}^2+\frac{1}{2}(M_2M_1)^2+\ldots)=0
\end{equation}

Only relevant terms are written explicitly. According to the rule of counting d.o.f, we conclude that
\begin{equation}
M_3**M_1**M_1**M_1+\frac{3}{2}M_2**M_2**M_1**M_1=0
\end{equation}

Note that (\ref{6gm2}) can be written as contractions of
\begin{equation}
\begin{split}
&\frac{1}{2}M_3**M_3+\frac{1}{6}M_2**M_2**M_2+\frac{1}{6\text{!}}\underbrace{M_1**M_1**\ldots**M_1}_{\# \ of\ M_1\ is\ 6}\\&+M_3**M_2**M_1+\frac{1}{6}M_3**M_1**M_1**M_1\\&+\frac{1}{4}M_2**M_2**M_1**M_1+\frac{1}{24}M_2**\underbrace{M_1**\ldots**M_1}_{\#\ of\ M_1\ is\ 4}=0
\end{split}
\end{equation}

Combining this with the two meson relations, we conclude that $M_2**M_2**M_2=0$.

\subsection{Quiver [3]-(3)-(3)-(3)-[3]}

Here we derive the 4th order meson relation of this quiver, which is $\frac{1}{2}M_2**M_2+M_1**M_3=0$.

\textit{Proof}:
As in the above proof, we write $M_2=dr_1l_1u=dQu$, $M_3=dr_1r_2l_2l_1u=dQ^2u$
\begin{equation}
\begin{split}
&\frac{1}{2}M_2**M_2+M_1**M_3\\
=&d_{[i}^{\alpha} \ d_{j]}^{\beta} \ u^{[k}_{\gamma}\ u^{l]}_{\rho}(\frac{1}{2}Q_{\alpha}^{\gamma}\ Q_{\beta}^{\rho}+\delta_{\alpha}^{\gamma}\ (Q^2)_{\beta}^{\rho})
\end{split}
\end{equation}

Therefore, we only need to prove that
\begin{equation}
\frac{1}{2}Q_{[\alpha}^{[\gamma}\ Q_{\beta]}^{\rho]}+\delta_{[\alpha}^{[\gamma}\ (Q^2)_{\beta]}^{\rho]}=0
\end{equation}

It's not very obvious, so we do a contraction:

\begin{equation}
Q^2-\text{Tr}(Q)Q+2Q^2-\text{Tr}(I)Q^2-\text{Tr}(Q^2)I
\end{equation} 
 
Where $I$ is a $3\times3$ identity matrix.

 So the vanishing of above equation follows from the fact that $\text{Tr}(I)=3$ and $\text{Tr}(Q^2)=0$.

\subsection{Higgs branch of $[SO(8)]-C_2-SO(2)$ }\label{higgs}
 
We consider the Higgs branch of quiver $[SO(8)]-C_2-SO(2)$. Using the F-term equations (\ref{ftm}), we can prove all the relations found in Coulomb branch of 2322.

2nd order
\begin{equation}
\begin{split}
\text{Tr}(M_1^2)&=Tr(A\Omega A^T A\Omega A^T)\\
&=\text{Tr}(BB^T\Omega BB^T\Omega)=0
\end{split}
\end{equation} 

3rd order
\begin{equation}
\begin{split}
M_1M_2&=A\Omega A^T A\Omega B\epsilon B^T\Omega A^T \\
&=A\Omega BB^T \Omega B\epsilon B^T\Omega A^T=0
\end{split}
\end{equation}

Similarly $M_2M_1=0$. And 
\begin{equation}
\epsilon _{i_1 i_2 j_1 j_2 j_3 j_4 j_5 j_6} (M_1)^{j_1 j_2} (M_1)^{j_3 j_4} (M_1)^{j_5 j_6}=0
\end{equation}
follows from $\text{rank}(M_1)\leq4$.

Finally,
\begin{equation}\label{eps}
\begin{split}
&\epsilon _{i_1 i_2 i_3 i_4 j_1 j_2 j_3 j_4} (M_1)^{j_1 j_2} (M_2)^{j_3 j_4}\\
=&\epsilon _{i_1 i_2 i_3 i_4 j_1 j_2 j_3 j_4}\Omega^{\alpha\beta}\Omega^{\gamma\theta}\Omega^{\sigma\rho}A_{j_1\alpha}A_{j_2\beta}\epsilon_{rs}A_{j_3\gamma}B_{\theta r}A_{j_4\rho}B_{\sigma s}\\
\end{split}
\end{equation}

Simplify the expression by writing 
\begin{equation}
\epsilon _{i_1 i_2 i_3 i_4 j_1 j_2 j_3 j_4}A_{j_1\alpha}A_{j_2\beta}A_{j_3\gamma}A_{j_4\rho}
\end{equation}
as
\begin{equation}
A_{i_1i_2i_3i_4}\epsilon_{\alpha\beta\gamma\rho}
\end{equation}

So equation (\ref{eps}) becomes
\begin{equation}
\begin{split}
&A_{i_1i_2i_3i_4}\epsilon_{\alpha\beta\gamma\rho}\Omega^{\alpha\beta}\Omega^{\gamma\theta}\Omega^{\sigma\rho}B_{\theta r}B_{\sigma s}\epsilon_{rs}\\
=2&A_{i_1i_2i_3i_4}\Omega^{\gamma\rho}\Omega^{\gamma\theta}\Omega^{\sigma\rho}B_{\theta r}B_{\sigma s}\epsilon_{rs}\\
=2&A_{i_1i_2i_3i_4}\Omega^{\sigma\theta}B_{\theta r}B_{\sigma s}\epsilon_{rs}=0
\end{split}
\end{equation}

4th order
\begin{equation}\label{exp1}
\begin{split}
&(M_1^2)_{['i_1[i_3}(M_1^2)_{i_2]'i_4]}\\
=&\Omega^{\alpha_1\beta_1}A_{['i_1\alpha_1}A_{k\beta_1}\Omega^{\beta_3\alpha_3}A_{k\beta_3}A_{[i_3\alpha_3}\Omega^{\alpha_2\beta_2}A_{i_2]'\alpha_2}A_{l\beta_2}\Omega^{\beta_4\alpha_4}A_{l\beta_4}A_{i_4]\alpha_4}\\
=&\Omega^{\alpha_1\beta_1}\Omega^{\alpha_2\beta_2}\Omega^{\alpha_3\beta_3}\Omega^{\alpha_4\beta_4}A_{[i_1\alpha_1}A_{i_2]\alpha_2}A_{[i_3\alpha_3}A_{i_4]\alpha_4}B_{\beta_1 r}B_{\beta_2 s}B_{\beta_3 r}B_{\beta_4 s}
\end{split}
\end{equation}

\begin{equation}\label{exp2}
\begin{split}
&(M_2)_{i_1i_2}(M_2)_{i_3i_4}\\
=&\Omega^{\alpha_1\beta_1}\Omega^{\beta_2\alpha_2}\epsilon_{rs}A_{i_1\alpha_1}B_{\beta_1 r}A_{i_2 \alpha_2}B_{\beta_2 s}\Omega^{\alpha_3\beta_3}\Omega^{\beta_4\alpha_4}\epsilon_{pq}A_{i_3\alpha_3}B_{\beta_3P}A_{i_4\alpha_4}B_{\beta_4 q}\\
=&\Omega^{\alpha_1\beta_1}\Omega^{\alpha_2\beta_2}\Omega^{\alpha_3\beta_3}\Omega^{\alpha_4\beta_4}A_{[i_1\alpha_1}A_{i_2]\alpha_2}A_{[i_3\alpha_3}A_{i_4]\alpha_4}\epsilon_{rs}\epsilon_{pq}B_{\beta_1 r}B_{\beta_2 s}B_{\beta_3 p}B_{\beta_4 q}
\end{split}
\end{equation}

It's easy to see that the two expressions (\ref{exp1}) and (\ref{exp2}) have the relation
\begin{equation}
2(M_1^2)_{['i_1[i_3}(M_1^2)_{i_2]'i_4]}=(M_2)_{i_1i_2}(M_2)_{i_3i_4}
\end{equation}

\section{Results of some $A_n$-type quiver}\label{some}
We list some of the results about $A_2$, $A_3$, $A_4$ and $A_5$ quivers in the end of this paper, including their PL, mirror and relations in each order.

\begin{sidewaystable}[h]
\begin{tabular}{|C{1cm}|C{11cm}|C{4.5cm}|C{3.5cm}|}
\hline
Quiver & PL & Mirror & Relation\\ \hline
11 & $ [11]t-(1+[11])t^2 $ & \input{11.tex} & \tabincell{c}{2nd:matrix} \\ \hline 
12 & $ [11]t-t^2-t^3 $ & \input{12.tex} & \tabincell{c}{2nd:trace\\3rd:trace} \\ \hline 
22 & $ [11]t+[11]t^2-(2+[11])t^3-(1+[11])t^4 $ & \input{22.tex} & \tabincell{c}{3rd:trace;matrix\\4th:matrix} \\ \hline 
23 & $ [11]t+[11]t^2-t^3-(2+[11])t^4 $ & \input{23.tex} & \tabincell{c}{3rd:trace\\4th:trace;matrix} \\ \hline 
24 & $ [11]t+[11]t^2-t^3-t^4-t^5-t^6 $ & \input{24.tex} & \tabincell{c}{3rd:trace\\4th:trace\\5th:trace\\6th:trace} \\ \hline 
33 & $ [11]t+[11]t^2+[11]t^3-(2+[11])t^4-(2+[11])t^5-(1+[11])t^6 $ & \input{33.tex} & \tabincell{c}{4th:trace;matrix\\5th:trace;matrix\\6th:matrix} \\ \hline 
\end{tabular}
\caption{$A_2$ quiver(1)}
\end{sidewaystable}

\begin{sidewaystable}
\begin{tabular}{|C{1cm}|C{11cm}|C{4.5cm}|C{3.5cm}|}
\hline
34 & $ [11]t+[11]t^2+[11]t^3-t^4-(2+[11])t^5-(2+[11])t^6 $ & \input{34.tex} & \tabincell{c}{4th:trace\\5th:trace;matrix\\6th:trace;matrix} \\ \hline 
35 & $ [11]t+[11]t^2+[11]t^3-t^4-t^5-(2+[11])t^6-t^7 $ & \input{35.tex} & \tabincell{c}{4th:trace\\5th:trace\\6th:trace;matrix\\7th:trace} \\ \hline 
36 & $ [11]t+[11]t^2+[11]t^3-t^4-t^5-t^6-t^7-t^8-t^9 $ & \input{36.tex} & \tabincell{c}{4th:trace\\5th:trace\\6th:trace\\7th:trace\\8th:trace\\9th:trace} \\ \hline 
44 & $ [11]t+[11]t^2+[11]t^3+[11]t^4-(2+[11])t^5-(2+[11])t^6-(2+[11])t^7-(1+[11])t^8 $ & \input{44.tex} & \tabincell{c}{5th:trace;matrix\\6th:trace;matrix\\7th:trace;matrix\\8th:matrix} \\ \hline 
45 & $ [11]t+[11]t^2+[11]t^3+[11]t^4-t^5-(2+[11])t^6-(2+[11])t^7-(2+[11])t^8 $ & \input{45.tex} & \tabincell{c}{5th:trace\\6th:trace;matrix\\7th:trace;matrix\\8th:trace;matrix} \\ \hline 
46 & $ [11]t+[11]t^2+[11]t^3+[11]t^4-t^5-t^6-(2+[11])t^7-(2+[11])t^8-t^9 $ & \input{46.tex} & \tabincell{c}{5th:trace\\6th:trace\\7th:trace;matrix\\8th:trace;matrix\\9th:trace} \\ \hline 
\end{tabular}
\caption{$A_2$ quiver(2)}
\end{sidewaystable}

\begin{sidewaystable}
\begin{tabular}{|C{1cm}|C{9cm}|C{5.2cm}|C{3.3cm}|}
\hline
47 & $ [11]t+[11]t^2+[11]t^3+[11]t^4-t^5-t^6-t^7-(2+[11])t^8-t^9-t^{10} $ & \input{47.tex} & \tabincell{c}{5th:trace\\6th:trace\\7th:trace\\8th:trace;matrix\\9th:trace} \\ \hline 
48 & $ [11]t+[11]t^2+[11]t^3+[11]t^4-t^5-t^6-t^7-t^8-t^9-t^{10}-t^{11}-t^{12} $ & \input{48.tex} & \tabincell{c}{5th-12th:trace} \\ \hline 
\end{tabular}
\caption{$A_2$ quiver(3)}
\end{sidewaystable}


\begin{sidewaystable}

\begin{tabular}{|C{1cm}|C{11cm}|C{4cm}|C{4cm}|}
\hline
Quiver & PL & Mirror & Relation\\ \hline
111 & $ [101]t -([101]+[020]+1)t^2 $ & \input{111.tex} & \tabincell{c}{2nd:M**M=0} \\ \hline 
121 & $ [101]t-([101]+1)t^2 $ & \input{121.tex} & \tabincell{c}{2nd:matrix} \\ \hline 
122 & $ [101]t-t^2-([101]+1)t^3 $ & \input{122.tex} & \tabincell{c}{2nd:trace\\3rd:matrix} \\ \hline 
123 & $ [101]t-t^2-t^3-t^4 $ & \input{123.tex} & \tabincell{c}{2nd:trace\\3rd:trace\\4th:trace} \\ \hline 
222 & $ [101]t+[101]t^2-(2[101]+[020]+2)t^3+(-[020]+[101]-1)t^4 $ & \input{222.tex} & \tabincell{c}{3rd:matrix;$ M_1**M_2=0 $\\4th:$ M_2**M_2=0 $} \\ \hline 
232 & $ [101]t+[101]t^2-([101]+2)t^3-([020]+2(101)+2)t^4 $ & \input{232.tex} & \tabincell{c}{3rd:trace;matrix\\4th:matrix;$ M_2**M_2=0 $} \\ \hline 
\end{tabular}
\caption{$A_3$ quiver(1)}
\end{sidewaystable}

\begin{sidewaystable}
\begin{tabular}{|C{1cm}|C{11cm}|C{4.5cm}|C{3.5cm}|}
\hline
233 & $ [101]t+[101]t^2-t^3-([020]+2[101]+3)t^4 $ & \input{233.tex} & \tabincell{c}{3rd:trace\\4th:trace;matrix;\\$M_2**M_2=0$} \\ \hline 
234 & $ [101]t+[101]t^2-t^3-([020]+[101]+2)t^4-t^5 $ & \input{234.tex} & \tabincell{c}{3rd:trace\\4th:trace;$ M_2**M_2=0 $\\5th:trace} \\ \hline 
242 & $ [101]t+[101]t^2-([101]+2)t^3-(1+[101])t^4 $ & \input{242.tex} & \tabincell{c}{3rd:trace;matrix\\4th:matrix} \\ \hline 
243 & $ [101]t+[101]t^2-t^3-(2+[101])t^4-(1+[101])t^5 $ & \input{243.tex} & \tabincell{c}{3rd:trace\\4th:matrix;trace\\5th:matrix} \\ \hline 
244 & $ [101]t+[101]t^2-t^3-t^4-([101]+2)t^5-([101]+1)t^6 $ & \input{244.tex} & \tabincell{c}{3rd: trace\\4th:trace\\5th:trace;matrix\\6th:matrix} \\ \hline 
245 & $ [101]t+[101]t^2-t^3-t^4-t^5-([101]+2)t^6 $ & \input{245.tex} & \tabincell{c}{3rd:trace\\4th:trace\\5th:trace\\6th:trace;matrix} \\ \hline 
\end{tabular}
\caption{$A_3$ quiver(2)}
\end{sidewaystable}

\begin{sidewaystable}
\begin{tabular}{|C{1cm}|C{11cm}|C{3.5cm}|C{4.5cm}|}
\hline
246 & $ [101]t+[101]t^2-t^3-t^4-t^5-t^6-t^7-t^8 $ & \input{246.tex} & \tabincell{c}{3rd:trace\\4th:trace\\5th:trace\\6th:trace\\7th:trace\\8th:trace} \\ \hline 
333 & $ [101]t+[101]t^2+[101]t^3-([020]+2[101]+3)t^4-([020]+2(101)+2)t^5+(2(101)-[020]+1)t^6 $ & \input{333.tex} & \tabincell{c}{4th:trace;matrix\\ $\frac{1}{2}M_2**M_2+M_1**M_3=0$;\\5th:matrix;$ M_2**M_3=0 $\\6th:$ M_3**M_3=0 $} \\ \hline 
343 & $ [101]t+[101]t^2+[101]t^3-([101]+2)t^4-(2(101)+[020]+3)t^5-(2(101)+[020]+2)t^6 $ & \input{343.tex} & \tabincell{c}{4th:trace;matrix\\5th:trace;matrix;\\$ M_2**M_3=0 $\\6th:matrix;$ M_3**M_3=0 $} \\ \hline 
344 & $ [101]t+[101]t^2+[101]t^3-t^4-([020]+2(101)+3)t^5-([020]+2(101)+3)t^6 $ & \input{344.tex} & \tabincell{c}{4th:trace\\5th:trace;matrix;\\$ M_2**M_3=0 $\\6th:trace;matrix;\\$M_3**M3=0$} \\ \hline 
345 & $ [101]t+[101]t^2+[101]t^3-t^4-([020]+[101]+2)t^5-([020]+2[101]+3)t^6-t^7 $& \input{345.tex} & \tabincell{c}{4th:trace\\5th:trace;$M_2**M_3=0$\\6th:trace;matrix;\\$M_3**M_3=0$\\7th:trace} \\ \hline
353 & $ [101]t+[101]t^2+[101]t^3-([101]+2)t^4-([101]+2)t^5-([020]+2[101]+2)t^6 $ & \input{353.tex} & \tabincell{c}{4th:trace;matrix\\5th:trace;matrix\\6th:matrix;$ M_3**M_3=0 $} \\ \hline 
\end{tabular}
\caption{$A_3$ Quiver(3)}
\end{sidewaystable}


\begin{sidewaystable}
\begin{tabular}{|C{1cm}|C{11cm}|C{4cm}|C{4cm}|}
\hline
1111 & $ [1001]t-([1001]+[0110]+1)t^2 $ & \input{1111.tex} & \tabincell{c}{2nd:$ M_1**M_1=0 $} \\ \hline 
1221 & $ [1001]t-([1001]+1)t^2 $ & \input{1221.tex} & \tabincell{c}{2nd:matrix} \\ \hline 
1222 & $ [1001]t-t^2-([0110]+[1001]+1)t^3 $ & \input{1222.tex} & \tabincell{c}{2nd:trace\\3rd:$ M_1**M_1**M_1=0 $} \\ \hline 
1232 & $ [1001]t-t^2-([1001]+1)t^3 $ & \input{1232.tex} & \tabincell{c}{2nd:trace\\3rd:matrix} \\ \hline 
1233 & $ [1001]t-t^2-t^3-([1001]+1)t^4 $ & \input{1233.tex} & \tabincell{c}{2nd:trace\\3rd:trace\\4th:matrix} \\ \hline 
1234 & $ [1001]t-t^2-t^3-t^4-t^5 $ & \input{1234.tex} & \tabincell{c}{2nd:trace\\3rd:trace\\4th:trace\\5th:trace} \\ \hline 
2222 & $ [1001] t+[1001] t^2-2([1001]+[0110]+1)t^3 $ & \input{2222.tex} & \tabincell{c}{3rd:$ M_1**M_2=0 $;\\$ M_1**M_1**M_1=0 $} \\ \hline 
2332 & $ [1001] t+[1001] t^2-([1001]+2)t^3-(2[0110]+3[1001]+2)t^4 $ & \input{2332.tex} & \tabincell{c}{3rd:trace;matrix\\4th:matrix;$ M_2**M_2=0 $;\\$M_1**M_1**M_2=0$} \\ \hline 
\end{tabular}
\caption{A4 Quiver}
\end{sidewaystable}

\begin{sidewaystable}
\begin{tabular}{|C{1cm}|C{11cm}|C{4cm}|C{4cm}|}
\hline
11111 & $ [10001] t-([01010]+[10001]+1) t^2 $ & \input{11111.tex} & \tabincell{c}{2nd:$ M_1**M_1=0 $;} \\ \hline 
12221 & $ [10001] t-([10001]+1)t^2-([00200]-[10001])t^3 $ & \input{12221.tex} & \tabincell{c}{2nd:matrix\\3rd:$ M_1**M_1**M_1=0 $} \\ \hline 
12222 & $ [10001] t-t^2-([00200]+[01010]+[10001]+1) t^3 $ & \input{12222.tex} & \tabincell{c}{2nd:trace;\\3rd:$ M_1**M_1**M_1=0 $} \\ \hline 
12321 & $ [10001] t-([10001]+1) t^2 $ & \input{12321.tex} & \tabincell{c}{2nd:matrix} \\ \hline 

\end{tabular}
\caption{A5 Quiver}
\end{sidewaystable}

\begin{sidewaystable}
	\centering
		\begin{tabular}{|c|c|c|c|c|c|c|c|c|}
		\hline
		Mirror theory&3&4&5&6&7&8&9&10 \\ \hline
		\tabincell{c}{\{2,4,1\}\\ \{0,5,0\}}&m&m; $M_2**M_2=0$&&&&&& \\ \hline
		\tabincell{c}{\{1,2,4,1\}\\ \{0,0,5,0\}}&tr&tr;m; $M_2**M_2=0$&m&&&&& \\ \hline
		\tabincell{c}{\{1,2,3,4,1\}\\ \{0,0,0,5,0\}}&tr&tr; $M_2**M_2=0$&tr;m&&&&& \\ \hline
		\tabincell{c}{\{1,2,3,4,5,1\}\\ \{0,0,0,0,5,0\}}&tr&tr; $M_2**M_2=0$&tr&tr&&&& \\ \hline
			
		\tabincell{c}{\{2,4,2\}\\ \{0,5,0\}}&tr; m&m&$M_2**M_2**M_1=0$&$M_2**M_2**M_2=0$&&&& \\ \hline
		\tabincell{c}{\{1,2,4,2\}\\ \{0,0,5,0\}}&tr&tr; m&m; $M_2**M_2**M_1=0$&$M_2**M_2**M_2=0$&&&& \\ \hline
		\tabincell{c}{\{1,2,3,4,2\}\\ \{0,0,0,5,0\}}&tr&tr&tr; m; $M_2**M_2**M_1=0$&$M_2**M_2**M_2=0$&&&& \\ \hline
		
		\tabincell{c}{\{1,3,5,2\}\\ \{0,0,5,0\}}&tr&tr;m&m&$M_2**M_2**M_2=0$&&&& \\ \hline
		\tabincell{c}{\{1,2,3,5,2\}\\ \{0,0,0,5,0\}}&tr&tr&tr; m&m; $M_2**M_2**M_2=0$&&&& \\ \hline
		\tabincell{c}{\{1,2,3,4,5,2\}\\ \{0,0,0,0,5,0\}}&tr&tr&tr&tr; m; $M_2**M_2**M_2=0$&&&& \\ \hline
		\tabincell{c}{\{1,2,3,4,5,6,2\}\\ \{0,0,0,0,0,5,0\}}&tr&tr&tr&tr;$M_2**M_2**M_2=0$&tr&&& \\ \hline
		
		\tabincell{c}{\{2,4,6,3\}\\ \{0,0,5,0\}}&tr&tr;m&m&&&&& \\ \hline
		\tabincell{c}{\{1,2,4,6,3\}\\ \{0,0,0,5,0\}}&tr&tr&tr;m&m&&&& \\ \hline
		\tabincell{c}{\{1,2,3,4,6,3\}\\ \{0,0,0,0,5,0\}}&tr&tr&tr&tr; m&m&&& \\ \hline\
		\tabincell{c}{\{1,2,3,4,5,6,3\}\\ \{0,0,0,0,0,5,0\}}&tr&tr&tr&tr&tr;m&m&& \\ \hline
		\tabincell{c}{\{1,2,3,4,5,6,7,3\}\\ \{0,0,0,0,0,0,5,0\}}&tr&tr&tr&tr&tr&tr; m&& \\ \hline
		\tabincell{c}{\{1,2,3,4,5,6,7,8,4\}\\ \{0,0,0,0,0,0,5,0\}}&tr&tr&tr&tr&tr&tr&tr&tr \\ \hline
		\end{tabular}
		\caption{Prediction of $A_4$ quiver}\label{pred}
\end{sidewaystable}
\bibliography{biblio}
\end{document}